\documentclass[aps,floats]{revtex4}
\usepackage{amsmath,amssymb}
\usepackage{graphicx,epsfig}

\begin{document}
\bibliographystyle {plain}

\def\oppropto{\mathop{\propto}} 
\def\opsimeq{\mathop{\simeq}}
\def\opoverderline{\mathop{\overline}}
\def\operarrow{\mathop{\longrightarrow}}
\def\opsim{\mathop{\sim}}

\def\fig#1#2{\includegraphics[height=#1]{#2}}
\def\figx#1#2{\includegraphics[width=#1]{#2}}


\title{ Driven interfaces in random media at finite temperature : \\
 is there an anomalous zero-velocity phase at small external force  ?
 } 


 \author{ C\'ecile Monthus and Thomas Garel }
  \affiliation{Institut de Physique Th\'{e}orique, CNRS and 
CEA Saclay, 91191 Gif-sur-Yvette cedex, France} 

\begin{abstract}

The motion of driven interfaces in random media at finite temperature $T$ and small external force $F$ is usually described by a linear displacement $h_G(t) \sim V(F,T) t$ at large times, where the velocity vanishes according to the creep formula as $V(F,T) \sim e^{-K(T)/F^{\mu}}$ for $F \to 0$. In this paper, we question this picture on the specific example of the directed polymer in a two dimensional random medium. We have recently shown (C. Monthus and T. Garel, arxiv:0802.2502) that its dynamics for $F=0$ can be analyzed in terms of a strong disorder renormalization procedure, where the distribution of renormalized barriers flows towards some "infinite disorder fixed point". In the present paper, we obtain that for small $F$, this "infinite disorder fixed point" becomes a "strong disorder fixed point" with an exponential distribution of renormalized barriers. The corresponding distribution of trapping times then only decays as a power-law $P(\tau) \sim 1/\tau^{1+\alpha}$, where the exponent $\alpha(F,T)$ vanishes as $\alpha(F,T) \propto F^{\mu}$ as $F \to 0$. Our conclusion is that in the small force region $\alpha(F,T)<1$, the divergence of the averaged trapping time $\overline{\tau}=+\infty$ induces strong non-self-averaging effects that invalidate the usual creep formula obtained by replacing all trapping times by the typical value. We find instead that the motion is only sub-linearly in time $h_G(t) \sim  t^{\alpha(F,T)}$, i.e. the asymptotic velocity vanishes $V=0$. This analysis is confirmed by numerical simulations of a directed polymer with a metric constraint driven in a traps landscape. We moreover obtain that the roughness exponent, which is governed by the equilibrium value $\zeta_{eq}=2/3$ up to some large scale, becomes equal to $\zeta=1$ at the largest scales.

\end{abstract}

\maketitle

\section{ Introduction   } 

\subsection{ Dynamical phase diagrams in the presence of quenched disorder  }

Transport phenomena in random media have remained a very active field
of research since the discovery of the Anderson localization 
fifty years ago \cite{anderson}.
For classical systems also, the presence of quenched disorder can induce 
completely new transport behaviors with respect to pure systems in some
regions of parameters.
It turns out that even the dynamics of a single particle in one-dimensional random media (see the review \cite{jpbreview} and references therein)
can already present a very rich phase diagram as a function 
of the temperature $T$ and the external applied force $F$.
As an example, we shown on Fig. 1 the exactly known phase diagram
 for the biased Sinai model 
\cite{kesten,derrida_pom,feigelman_vin,jpb_annphys} :
one needs to introduce a dimensionless parameter $\alpha(F,T)= T F$
(where we have chosen to fix the disorder strength at some simple value
to simplify the notations and emphasize the dependence upon $T$ and $F$
we are interested in ).
For $F=0$, the motion is logarithmically slow $x(t) \sim \pm (\ln t )^2$.
For $0<\alpha(F,T)<1$, the diffusion is anomalous with a sub-linear displacement $x \sim t^{\alpha(F,T)}$  ; for $1<\alpha<2$, the velocity becomes finite $x \sim V t$, but the dispersion remains anomalous of order $t^{1/\alpha(F,T)}$.
Finally for $\alpha>2$, both the velocity and the diffusion coefficient 
are finite. So already in simple one-dimensional systems, 
various dynamical properties can undergo phase transitions at various thresholds.

\begin{figure}[htbp]
\includegraphics[height=8cm]{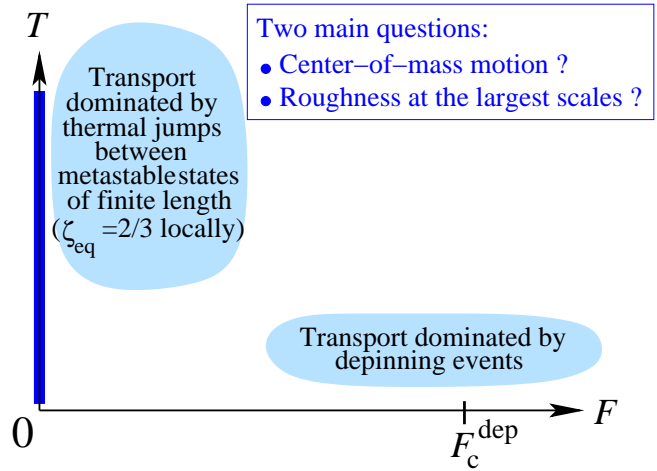}
\caption{ Exactly known dynamical phase diagram 
as a function of the temperature $T$
and the external force $F$ of
 the one-dimensional Sinai model \cite{jpbreview} : 
the important dimensionless parameter
is $\alpha(F,T)=T F$ (the disorder strength has been
fixed at some simple value). From the point of view of strong disorder 
renormalization \cite{sinairg,sinaibiasdirectedtraprg,review}, the  
logarithmic behavior $ x \sim \pm (\ln t)^2 $
for $F=0$ corresponds to an ``infinite disorder fixed point'',
whereas the anomalous diffusion phase $x \sim t^{\alpha}$
for $0<\alpha<1$ corresponds to a ``strong disorder fixed point''.
A finite-velocity $V>0$ appears only for $\alpha>1$
 , i.e. when the force is above some temperature-dependent value.
 }
\label{figsinai}
\end{figure}

\begin{figure}[htbp]
\includegraphics[height=8cm]{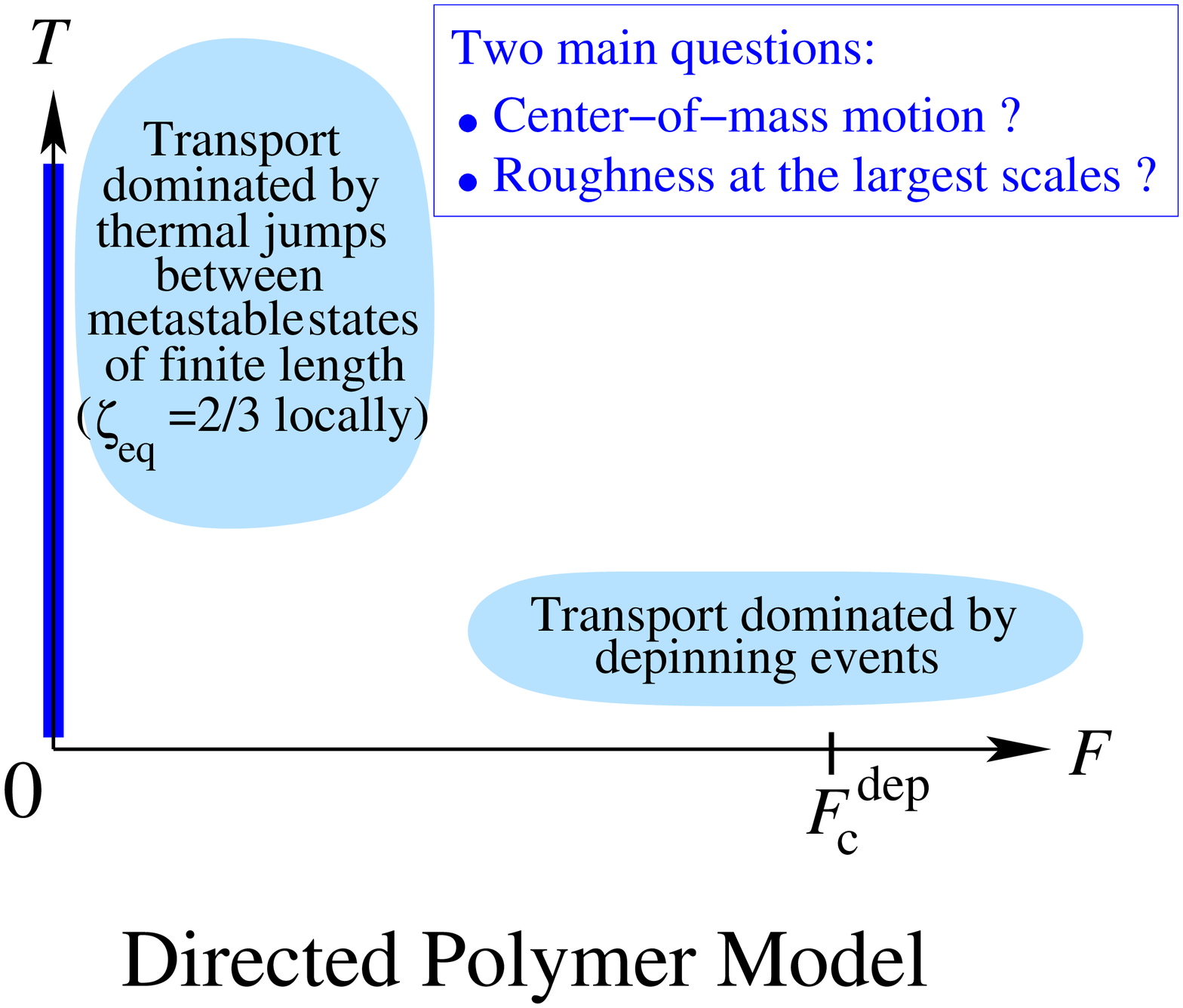}
\caption{ The dynamical phase diagram for the directed polymer
 in a two-dimensional random medium as a function of the temperature $T$
and the external force $F$ is not exactly known. 
However one expects physically two very different regimes :
(i) near the zero-temperature depinning critical point $F_c^{dep}$,
the transport at small temperature is dominated by depinning events.
(ii) in the regime where $T$ is finite and $F$ is small,
the roughness exponent $\zeta_{eq}=2/3$ of the equilibrium case $F=0$
is expected to describe the polymer roughness up to some large
scale (that diverges as $F \to 0$). The transport is then dominated
by thermal jumps of finite segments of the polymer
between quasi-equilibrated metastable states. 
In this paper, we focus only on the regime (ii) and discuss
the center-of-mass motion and the roughness at the largest scales. }
\label{figdp}
\end{figure}

The case of more complex systems like interfaces or manifolds in random media
has attracted a lot of attention in relation with many important 
applications, and various approaches have been developed to
elucidate the structure of the dynamical phase diagram
(see the reviews \cite{dsfreview,TGreview} and references therein).
The simplest model within this class of systems is
 the directed polymer in a two-dimensional random medium 
(see \cite{Hal_Zha} for a review), first
 introduced to model an interface
 in the low-temperature phase of two-dimensional
disordered ferromagnets \cite{Hus_Hen}. 
The statics of this model is rather well understood with
exactly known critical exponents :
at any temperature $T$, the directed polymer
 is in a disordered-dominated phase characterized
by the roughness exponent $\zeta_{eq}=2/3$ and
 the droplet exponent $\theta=2 \zeta_{eq}-1=1/3$ \cite{Hal_Zha}.
In comparison, the dynamics is much harder to study,
  and no exact result exists.
Even numerically, the computational complexity changes completely
between the statics that can be studied via exact transfer matrix methods
that are polynomial in the size of the polymer \cite{Hal_Zha},
and the dynamics where the determination of barriers is an NP-complete
problem \cite{middleton}. As a consequence,
a complete characterization of the dynamical properties as a 
function of the temperature and external force 
(here we consider that the disorder strength is fixed)
 has remained a very challenging issue.
An essential novelty with respect
 to the dynamics of the Sinai model described above,
 is the presence of the chain constraint of the polymer,
so that even at zero temperature $T=0$, the dynamics is already non trivial :
the competition between interaction and disorder gives rise
to a non-equilibrium depinning phase transition, 
between a pinned phase $F<F_c^{dep}$
where the interface remains blocked forever in some configuration, and 
a moving phase for $F>F_c^{dep}$.
This type of depinning transition has motivated a lot of experimental
 and theoretical studies (see the reviews \cite{dsfreview,TGreview}),
and an exact mean-field solution has been obtained in \cite{Der_Van}.
The critical region with a small temperature 
around the depinning transition of an elastic line
 has been studied in \cite{rounding}. 

In the present paper, we focus on the opposite region of the dynamical phase diagram
(see Fig. 2) where the temperature is finite and the external force is small.
In this regime, one expects that the dynamics is dominated
 by thermal activations between locally quasi-equilibrated metastable states.
In particular, the polymer is expected to keep during the motion
its equilibrium roughness exponent $\zeta_{eq}=2/3$ up to some large
scale (that diverges as $F \to 0$). This equilibrium roughness
 has been measured
in dynamic simulations in \cite{kaper}, and in
\cite{Kol_tworegimes} when the temperature remains 
larger that some disorder-dependent threshold. 
However in other regimes, larger roughness exponents 
have been measured at larger scales  \cite{Kol_tworegimes,Kol_below}.
This issue of the roughness at the largest scales will be rediscussed
later in the paper. But we first need to better 
 understand the statistical properties of the barriers
 between the locally quasi-equilibrated metastable states.

\subsection{ Usual creep scenario for
finite temperature and small external force }

In the regime of finite temperature and small external force,  
the standard picture in the field 
(see the review \cite{TGreview} and references therein) seems to be
 that the dynamics then corresponds to a 'creep' motion, 
where the center-of-mass moves linearly in time
 as soon as the temperature is positive 
\begin{eqnarray}
h_G(t) \opsimeq_{t \to \infty} V(F,T) \ t
\label{defvitesse}
\end{eqnarray}
with a very non-linear force-velocity relation of the form
\begin{eqnarray}
V(F,T) \opsimeq_{F \to 0} e^{ - \frac{K(T)}{F^{\mu} } } 
\label{vitessecreep}
\end{eqnarray}
where the exponent $\mu$ involves the dimension $d$ of the interface
(the directed polymer we are interested in corresponds to $d=1$),
the equilibrium roughness exponent $\zeta_{eq}$, and the barrier exponent $\psi$
of the dynamics without external force
\begin{eqnarray}
\mu = \frac{\psi}{(\zeta_{eq} +d-\psi)}
\label{muusual}
\end{eqnarray}
Note that this relation is usually written 
in the form $\mu_{usual}=(d-2+2 \zeta_{eq})/(2-\zeta_{eq})$
after using the additional assumption 
concerning the identity between the barrier exponent $\psi$ 
and the droplet exponent $\theta=2 \zeta_{eq} +d-2$. However, in our opinion,
 there is no convincing evidence of the equality $\psi=\theta$
neither theoretically 
(see \cite{conjecturepsi} for a recent summary on the debate)
nor numerically 
(the best numerical results presently available \cite{rosso} points towards
an exponent of order $\psi \sim 0.49$ rather different from the 
exactly known droplet exponent $\theta=1/3$), 
 we will consider in the following
 that the barrier $\psi$ is an independent exponent
satisfying the bound
\begin{eqnarray}
\psi \geq \theta
\label{psibound}
\end{eqnarray}

\subsection{ Summary of the present paper : 
alternative scenario with a zero-velocity phase at small
external force }

The usual qualitative argument given in favor of a finite velocity 
in Eq. \ref{defvitesse} as soon as $T>0$ is that 
'any barrier can be passed by thermal activation'.
This is of course true, but this is not sufficient 
to conclude that the asymptotic velocity is finite,
since the motion can also exhibit anomalous behavior
as shown by the Sinai model at small external force
 (see the beginning of the introduction).
To determine the asymptotic motion of the center-of mass
at large times, one needs to study the time needed
 to travel over a given large distance.
The more quantitative argument in favor of the finite velocity
of Eq. \ref{defvitesse} is a scaling argument where the barrier 
landscape existing in the absence of bias
is 'tilted' to take into account the bias contribution : 
this scaling argument yields that the relevant barriers
for the large-time dynamics corresponds to a barrier scale 
$B^*(F,T)$ and to a length scale $l^*(F,T)$ that are finite
as soon as $F>0$ (and that diverge as $F \to 0$). 
We fully agree with this scaling argument that will be rediscussed 
below in section \ref{scalingarg}. However again, 
this is not sufficient to conclude that the velocity is finite,
since exactly the same argument can be made for the Sinai model
 where an anomalous diffusion phase exists
(see section \ref{scalingarg} for more details).
The crucial property to determine 
whether the velocity is finite or not is the probability distribution of
barriers $B$ around this typical value $B^*(F,T)$.
In the present paper, we explain 
that within the strong disorder renormalization
in configuration space that we have introduced recently  \cite{letter}
(see \cite{rgrateslong} for a more detailed presentation), 
one obtains an exponential tail for
the distribution of renormalized barriers
\begin{eqnarray}
P(B) \oppropto_{B \to +\infty} \ e^{ -\frac{B }{B^*(F,T)} }
\label{distriexp}
\end{eqnarray}
As is well known in the field of disordered systems
 since Derrida's Random Energy Model \cite{rem},
other disordered models sharing the same low-energy states 
statistics \cite{jpb_extreme,glatz}
and related Bouchaud's trap Models \cite{jpb_trap}
(see also \cite{rammal} for one of the first mention of exponentially distributed barriers in connection with extremal statistics),
this 'innocent' exponential distribution for the barriers
 corresponds via the change of variable
$\tau=e^B$ to a very broad power-law decay for
the distribution of the trapping times
 $\tau$
\begin{eqnarray}
P(\tau) \oppropto_{\tau \to +\infty}
 \frac{ \alpha(F,T) }{\tau^{1+\alpha(F,T)} } 
\label{powerlaw}
\end{eqnarray}
The exponent depends continuously 
on the external force and on the temperature
(we consider here that the disorder remains fixed)
\begin{eqnarray}
 \alpha(F,T)= \frac{1}{B^*(F,T)} 
\label{alphaft}
\end{eqnarray}
Since the characteristic barrier scale $B^*(F,T)$
 grows with $F$ and diverges as $F \to 0$ as
\begin{eqnarray}
 B^*(F,T) \oppropto_{F \to 0} \frac{1}{ F^{\mu}}
\label{bfmu}
\end{eqnarray}
the exponent $\alpha(F,T)$ vanishes as $F \to 0$ as
\begin{eqnarray}
 \alpha(F,T) \oppropto_{F \to 0}  F^{\mu}
\label{alphafmu}
\end{eqnarray}
As a consequence, the region of small external force where $ \alpha(F,T) <1$
corresponds to a very broad distribution of trapping times
with a diverging averaged value $\overline{\tau} =+\infty$. 
This invalidates the usual creep formula that is obtained by replacing
all trapping times by the typical value $\tau_{typ} \sim e^{B_{typ}} 
\sim e^{B^*(F,T)}$ to obtain $V \sim 1/\tau_{typ}$
 (see Eq. \ref{vitessecreep}).
We obtain instead that for $ \alpha(F,T) <1$, the center-of-mass displacement then grows
 only sub-linearly in time
\begin{eqnarray}
h_G(t) \opsimeq_{t \to \infty}  t^{\alpha(F,T)}
\label{sublinear}
\end{eqnarray}
i.e. the asymptotic velocity vanishes $V=0$.

Although the vast majority of papers on the subject never
 mentions the possibility of a zero-velocity phase, 
we are aware of three papers where the question of the probability
distribution of barriers around the typical value has been raised :

(i) twenty years ago, Ioffe and Vinokur \cite{Iof_Vin}  
have proposed the anomalous sub-linear motion of Eq. \ref{powerlaw}.
But for reasons that are very unclear to us, 
 this possibility seems to have completely disappeared in the 
more recent literature.

(ii) fifteen years ago, Bouchaud has pointed out in  \cite{jpb_cargese}
that the usual creep argument might lead to incorrect results
because of problem of strong fluctuations, and cites the Sinai model as an example where
the transport is indeed harder than anticipated from the typical barrier alone.

(ii) ten years ago, Vinokur, Marchetti and Chen \cite{Vin_Mar} have proposed
an exponential distribution of barriers, based on 
 extremal statistics argument,
and the corresponding power-law distribution for trapping times.
However, these authors have supplemented this power-law distribution
by a sharp cut-off to recover the usual finite velocity behavior of 
Eq. \ref{defvitesse}. In our opinion, there is no good reason
to impose this sharp cut-off, since the same procedure for the
exactly soluble Sinai model would give a wrong answer.

\subsection{ Organization of the paper }

To summarize this long introduction, 
the aim of the present paper is 
 to justify the existence of the anomalous diffusion phase of Eq. \ref{sublinear}
in the region of the phase diagram 
corresponding to finite temperature and small external force (see Fig. 2),
via the use of some strong disorder renormalization procedure on the barriers.
The paper is organized as follows.
In section \ref{infinite}, we briefly recall the main idea of
strong disorder renormalization, and we describe the statistical properties
of the ``infinite disorder'' fixed point that describes the thermal dynamics
of the directed polymer without external force.
In section \ref{strong}, we explain how the presence of a small external force
$F$ transforms this "infinite disorder fixed point" into
a "strong disorder fixed point" with an exponential distribution
of renormalized barriers, and we describe the consequences for 
distribution of trapping times and the large-time
dynamics. In Section \ref{numerics}, we present detailed numerical simulations
for a directed polymer driven in a traps landscape.
In Section \ref{previous}, we compare our results concerning the
roughness with previous works.
Our conclusions are summarized in section \ref{conclusion}.
In Appendix A, we recall the subtleties associated to the type of
the interactions (metric constraint versus elastic energy) to justify
our choice to consider the metric constraint.

\section{ Properties of the 'infinite disorder fixed point' for $F=0$}

\label{infinite}

\subsection{ Strong disorder renormalization in configuration space   }

Strong disorder renormalization 
(see \cite{review} for a review) is a very specific type of renormalization
(RG) that has appeared in the field of quantum spin chains :
this approach introduced by Ma and Dasgupta \cite{madasgupta} 
has been developed by D.S. Fisher \cite{dsf}, who has introduced
the crucial notion of ``infinite disorder'' fixed points
where the method becomes asymptotically exact, and who has shown
how to obtain explicit exact results for
critical exponents and scaling functions for zero-temperature 
quantum critical points.
This method has thus generated a lot of activity for various
disordered quantum models \cite{review}.
It has been then successfully applied to
various classical disordered dynamical models,
such as random walks in random media \cite{sinairg,
sinaibiasdirectedtraprg},
reaction-diffusion in a random medium \cite{readiffrg}, 
coarsening dynamics of classical spin chains \cite{rfimrg}, 
trap models \cite{traprg}, 
random vibrational networks \cite{vibrational}
absorbing state phase transitions \cite{contactrg},
zero range processes \cite{zerorangerg} and 
exclusion processes  \cite{exclusionrg}.
In all these cases, the strong disorder RG rules 
have been formulated {\it in real space},
with specific rules depending on the problem.
For more complex systems where
 the formulation of strong disorder RG rules
has not been possible in real space, 
we have recently proposed in \cite{letter} a strong disorder 
RG procedure { \it in configuration space} that can be
 defined for any master equation as we now recall.

The starting point of strong disorder renormalization in configuration space
is the
 Master Equation describing the evolution of the
probability $P_t ({\cal C} ) $ to be in a configuration ${\cal C}$ at time t
\begin{eqnarray}
\frac{ dP_t \left({\cal C} \right) }{dt}
= \sum_{\cal C '} P_t \left({\cal C}' \right)
 W \left({\cal C}' \to  {\cal C}  \right) 
 -  P_t \left({\cal C} \right) W_{out} \left( {\cal C} \right)
\label{master}
\end{eqnarray}
The notation  
$ W \left({\cal C}' \to  {\cal C}  \right) $ 
represents the transition rate from configuration 
${\cal C}'$ to ${\cal C}$, and the notation
\begin{eqnarray}
W_{out} \left( {\cal C} \right)  \equiv
 \sum_{ {\cal C} '} W \left({\cal C} \to  {\cal C}' \right) 
\label{wcout}
\end{eqnarray}
represents the total exit rate out of configuration ${\cal C}$.

For dynamical models, the aim of any renormalization procedure
is to integrate over 'fast ' processes to obtain effective properties 
of 'slow' processes.
 The general idea of 'strong renormalization' for dynamical models
consists in eliminating iteratively the 'fastest' process.
The RG procedure introduced
in \cite{letter} consists in the
iterative elimination of the state with the highest exit rate.
We refer to \cite{letter,rgrateslong} for a detailed description and
derivation of the renormalization rules on the transition rates.
Their most important property is their multiplicative structure 
that suggests that for a very broad class of disordered systems,
 the distribution of renormalized exit barriers 
\begin{eqnarray}
B_{out} \equiv - \ln W_{out}
\label{wcrenormalizedout}
\end{eqnarray}
 will become broader and broader upon iteration, 
so that the strong disorder renormalization procedure
should become asymptotically exact at large time scales.
Note that a very important advantage of this formulation in terms
of the transition rates of the master equation is that 
the renormalized barriers take into account the true 'barriers'
of the dynamics, whatever their origin which can be
 either energetic or entropic.

After this general presentation (see \cite{letter,rgrateslong}
for more details), we now turn to the specific problem of
a directed polymer in a two-dimensional random medium
we are interested in.
For the dynamics { \it without external force}, we have
followed numerically 
the RG flow of the renormalized transition rates
and we have found some ``infinite disorder'' fixed point 
\cite{letter,rgrateslong} as we now explain.

\subsection{ Distribution of renormalized exit barriers at large scale}

The RG scale $\Gamma$ is defined as the scale of the last eliminated
exit barrier.
At time $t$, the appropriate RG scale is thus
\begin{eqnarray}
\Gamma(t)= \ln t
\end{eqnarray}
so that all metastable states of exit barrier $B_{out}>\Gamma$,
i.e. of exit time $\tau=e^{B_{out}} >t $ have been kept,
whereas all metastable states of exit time $\tau=e^{B_{out}} <t $
have been eliminated.
At large scale,
one expects that the probability distribution
 of the remaining exit barriers $B_{out} \geq \Gamma$ will 
converge towards some scaling form
\begin{eqnarray}
P_{\Gamma} ( B_{out}  ) \opsimeq_{ \Gamma \to \infty} 
  \frac{1}{\sigma(\Gamma) } {\hat P} 
\left( \frac{B_{out} - \Gamma}{\sigma(\Gamma) } \right)
\label{pgammabout}
\end{eqnarray}
where ${\hat P} $ is the fixed point probability distribution, 
and where $\sigma(\Gamma)$ is the appropriate scaling factor
that represents the width of the renormalized distribution.

For the directed polymer in a two-dimensional random medium,
we have obtained numerically in \cite{letter,rgrateslong} that 
the width $\sigma(\Gamma)$
grows asymptotically linearly with the RG scale $\Gamma$
\begin{eqnarray}
\sigma(\Gamma) \opsimeq_{ \Gamma \to \infty} \Gamma
\label{sigmalinear2}
\end{eqnarray}
and that the rescaled distribution of Eq. \ref{pgammabout} is 
extremely close to the exponential form
\begin{eqnarray}
{\tilde P} (x) \simeq e^{-x}
\label{pexp}
\end{eqnarray}
Note that these two properties
 seem extremely robust within strong disorder RG
since they hold for exactly in soluble models in $d=1$  \cite{review}
and have been also found numerically in quantum models 
in dimension $d>1$ \cite{motrunich}.

\subsection{ Scaling between barriers and length scales} 

To describe the thermal relaxation of disordered systems towards equilibrium 
when starting at time $t=0$ from a non-equilibrium initial state,
 it is useful to introduce
the notion of some coherence length $l_T(t)$ that grows slowly in time
(see the reviews \cite{bouchaud,berthierhouches}
 and references therein).
This coherence length
 separates the smaller lengths $l < l_T(t)$ which are quasi-equilibrated
at time $t$ from the larger lengths $l > l_T(t)$ which are completely
out of equilibrium. Then full equilibrium is reached only when
the coherence length reaches the macroscopic linear
size $l_T(t_{eq}) = L$ of the system. 
Within the droplet scaling theory
proposed both for spin-glasses \cite{heidelberg,Fis_Hus} 
and for directed polymers in 
random media \cite{Fis_Hus_DP}, the barriers grow 
as a power law of the length $l$
\begin{eqnarray}
B(l) \sim l^{\psi}
\label{defpsidrop}
\end{eqnarray}
with some barrier exponent $\psi>0$.
The typical time $t_{typ}(L)$ associated to scale $l$ grows as an exponential
$\ln t_{typ}(l) \sim B(l) \sim l^{\psi}$. 
Equivalently, the coherence
length-scale $l_T(t)$ associated to time $t$ grows only logarithmically in time
\begin{eqnarray}
l_T(t) \sim \left( \ln t \right)^{\frac{1}{\psi}}
\label{typltime}
\end{eqnarray}
In the numerical study of the relaxation towards equilibrium
of an elastic chain in a two dimensional random medium,
starting from a straight line at $t=0$, this coherence length can be extracted
from the behavior of the structure factor as a function of time \cite{rosso},
and the corresponding measure of the barrier exponent
yields $\psi \sim 0.49$  \cite{rosso} (see the comments on this exponent $\psi$
before Eq. \ref{psibound}).

Within the strong disorder renormalization procedure,
one may either extract a coherence length $l_{\Gamma}$ as a function
of the RG scale \cite{rgrateslong}, or consider
the statistics of the barriers associated to a given length 
\cite{letter,rgrateslong}, and one obtains the scaling
\begin{eqnarray}
B(l) \opsimeq_{l \gg 1}  \Delta(T) l^{\psi} u
\label{barrierpsi}
\end{eqnarray}
where $u$ is a random variable of order one.
The barrier exponent $\psi$ obtained in \cite{rgrateslong}
is of order $\psi \sim 0.47$, i.e. close to 
the estimation $\psi \sim 0.49$ measured in \cite{rosso} via Langevin
dynamics. A systematic study of the dependence of the prefactor
$\Delta(T)$ upon temperature is not yet available.
This prefactor would be of the form $C/T$ if the barriers were 
purely energetic, but since the strong disorder renormalization
is defined on the transition rates,
the prefactor $\Delta(T)$ may also contain entropic contributions
and then have a more complicated temperature dependence.

\subsection{ Final picture for the dynamics at $F=0$} 

The final picture for the dynamics of the directed polymer of length $L$
 at finite temperature $T$ and no external force $F=0$
 is thus the following :

(a) during the regime $l_T(t)  <L$ where the coherence length $l_T(t)$
is smaller than the length $L$ of the polymer, 
the polymer is quasi-equilibrated with its equilibrium roughness exponent 
$\zeta_{eq}=2/3$ only on smaller length scale
 than the coherence length $l<l_T(t)$,
whereas larger length scale $l>l_T(t)$ are still completely out-of-equilibrium.
The ``infinite disorder fixed point'' describes the hierarchical structure of
growing barriers that have to be passed
 to equilibrate on larger and larger length scales, and this is why
the coherence length grows only logarithmically in time
\begin{eqnarray}
l_T(t) \sim \left( \frac{\ln t}{\Delta(T)}\right)^{\frac{1}{\psi}}
\label{coherence}
\end{eqnarray}

(b)  when the coherence length $l_T(t)$ reaches the length $L$ of the polymer,
the full polymer is characterized by the
equilibrium roughness exponent $\zeta_{eq}=2/3$.
The coherence length $l_T(t)$ cannot grow anymore, so that the renormalization
procedure has to be stopped at the RG scale of order 
\begin{eqnarray}
\Gamma_L \sim \Delta(T) L^{\psi}
\label{gammaLstop}
\end{eqnarray} 
One then expects that the polymer of roughness exponent $\zeta_{eq}=2/3$
will become able to move slowly in the transversal direction between metastable
states that are separated by distances of order $L^{\zeta_{eq}}$
and by barriers distributed with the fixed point distribution 
( Eq. \ref{pgammabout}, \ref{sigmalinear2} and \ref{pexp})
for the scale $\Gamma_L$,  that presents the exponential decay
\begin{eqnarray}
P_{\Gamma_L} ( B_{out}  ) \opsimeq_{B_{out} \to \infty }
  \frac{1}{\Gamma_L } e^{- \frac{B_{out} }{\Gamma_L } }
\label{pgammafin}
\end{eqnarray}
On the renormalized scale $\Gamma_L$, one expects that
the transverse motion of the center-of-mass corresponds to  
an effective one-dimensional Sinai model, where the unit distance
scale is of order $L^{\zeta_{eq}}$ and the unit barrier scale is 
$\Gamma_L \sim \Delta(T) L^{\psi}$.
As a consequence for $\ln t \gg \Delta(T) L^{\psi}$, one expects 
the logarithmically slow behavior
\begin{eqnarray}
h_G(t) \simeq \pm L^{\zeta_{eq}}
 \left( \frac{ \ln t }{ \Delta(T) L^{\psi} } \right)^2
\label{sinaieffective}
\end{eqnarray}

\section{ Properties of the
'Strong disorder fixed point' for small force  $F$} 

\label{strong}

\subsection{ Analysis of the RG flow for small $F$ } 

The "infinite disorder fixed point" described above
 for $F=0$ characterizes some 'criticality' in the time direction
in the following sense : there is no 
characteristic scale for the barriers except the RG scale $\Gamma$ itself.
This scale invariance will be broken by
 the introduction of some external force $F$
that will introduce some characteristic length-scale
 and thus some corresponding scale $B^*(F,T)$ for the barriers.
However, if the external force is very small, 
the scale $B^*(F,T)$ (that diverges as $F \to 0$)
will be very large.
We thus expect that the RG flow can be analyzed in terms of two regimes :

(i) during the first regime $1 \ll \Gamma \leq  B^*(F,T)$, 
the distribution of the renormalized exit barriers
will follow the same RG flow as in the absence of force, 
i.e. it will converge towards the scaling of Eq. \ref{pgammabout},
with the same exponential rescaled distribution of Eq. \ref{pexp}
\begin{eqnarray}
P_{\Gamma,F,T} ( B_{out}  ) \opsimeq 
 \frac{1}{\sigma(\Gamma,F,T) } 
e^{ -  \frac{(B_{out} - \Gamma)}{ \sigma(\Gamma,F,T)}}
\label{pgammtransi}
\end{eqnarray}
with a width $\sigma(\Gamma,F,T)$
which grows linearly in $\Gamma$ as in Eq. \ref{sigmalinear2}
\begin{eqnarray}
\sigma(\Gamma,F,T) \opsimeq_{ 1 \ll \Gamma \ll  B^*(F,T) }  \Gamma
\label{sigmagammatransi}
\end{eqnarray}

(ii) when the large scale $\Gamma \sim B^*(F,T)$ is reached,
the width saturates
at the finite large value $B^*(F,T)$
\begin{eqnarray}
\sigma(\Gamma,F,T) \opsimeq_{ \Gamma \geq B^*(F,T)} B^*(F,T)
\label{sigmasaturation}
\end{eqnarray}
instead of the flow towards infinity that characterizes
 the critical case $F=0$.
Since this width remains finite asymptotically as $\Gamma \to \infty$, 
one speaks of a `finite-disorder fixed point'.
However, since this width $B^*(F,T)$ diverges at small force ,
the region of small $F$  is a  'strong disorder fixed point',
where the asymptotic accuracy of the 
renormalization approach is of order $1/B^*(F,T)$.
This notion is thus very useful to study the vicinity of 
'infinite disorder fixed point' in the space of parameters 
and we refer to the review \cite{review} for more detailed discussions.
In one-dimension, where the strong disorder RG procedure
can be followed exactly, the crossover of the width between the regimes
(i) and (ii) described above, is of the form
(translated in our present notation) : $\sigma(\Gamma,F,T)=B^*(F,T)
\left[ 1- e^{- \frac{\Gamma}{B^*(F,T)}} \right] $ \cite{dsf,review}.

For our present analysis, the important point is that 
the saturation scale $B^*(F,T)$ is large enough, so that the RG flow
during the first regime $1 \ll \Gamma \ll B^*(F,T)$,
 that behaves as the critical flow, contains
sufficiently RG steps to have converged towards the scaling form 
of Eq. \ref{pgammtransi}. So, 
when saturation occurs at scale $B^*(F,T)$, the probability
distribution of renormalized exit barrier 
follows the exponential form 
\begin{eqnarray}
P_{\Gamma  }(B) 
\opsimeq \frac{1} {B^*(F,T)} \ e^{ -\frac{(B - \Gamma)}{B^*(F,T)} }
\label{distriexp2}
\end{eqnarray}

\subsection{ Physical meaning of the saturation } 

The physical meaning of the two regimes described above is as follows.
During the first regime (i), the external force $F$ is so small
that the flow is very similar to the flow for $F=0$.
In particular, the barriers grows upon iteration
on scales $1 \ll \Gamma \ll  B^*(F,T)$
and the motion is not yet directed along the bias.
When the saturation occurs at the large scale $B^*(F,T)$,
this means on the contrary that the motion becomes effectively
directed in the direction of the external force $F$ for scales
$\Gamma>B^*(F,T)$. The barriers against the bias are not renormalized
anymore, and one can stop the renormalization procedure.
The appropriate model on this scale is then a directed model along
the bias, with barriers distributed as in Eq. \ref{distriexp2}.
As recalled in the introduction (see the discussion
between Eqs. \ref{distriexp} and \ref{powerlaw}), the
corresponding distribution of the trapping time $\tau=e^{B}$
is then a broad power law
\begin{eqnarray}
P(\tau) \opsimeq \frac{ 1 }{\tau^{1+\alpha(F,T)} } 
\label{powerlaw2}
\end{eqnarray}
with exponent
\begin{eqnarray}
 \alpha(F,T)= \frac{1}{B^*(F,T)} 
\label{alphaft2}
\end{eqnarray}

\subsection{ Determination of the saturation scale $B^*(F,T)$ } 

\label{scalingarg}

Within our present RG framework, the saturation scale $B^*(F,T)$
should be determined as the limiting value of the width 
(Eq. \ref{sigmasaturation}) of the renormalized
distribution of barriers.
To determine the dependence  on the external force at small $F$,
we may rephrase the scaling argument which is usually used in the field
\cite{TGreview} as follows.
On the length scale $l$, the barriers {\it in the absence of external
force} follow the scaling of Eq. \ref{barrierpsi}
with the barrier exponent $\psi$. If the external force $F$ is small,
one may take into account its effects by a 'tilt' of the landscape
that lowers the barriers against the force in the following way
\begin{eqnarray}
B(l) \opsimeq_{l \gg 1} 
 \Delta(T) l^{\psi} u - \frac{F}{T} l^{1+\zeta_{eq}}
\label{barrierpsidrift}
\end{eqnarray}
where $\zeta_{eq}$ represents the equilibrium roughness exponent,
so that the correction in Eq. \ref{barrierpsidrift} corresponds
to a transversal move of order $l^{\zeta_{eq}}$ for the segment
of length $l$ of the polymer.
Since $(1+\zeta_{eq})>\psi$, the force term always dominates at sufficiently
 large length scale. The length scale $l^*(F,T)$ that will give rise to
the biggest barriers can be obtained by differentiating Eq. 
\ref{barrierpsidrift} with respect to $l$. Dropping constants of order
$O(1)$ one obtains the length scale
\begin{eqnarray}
l^*(F,T) \opsimeq_{F \to 0} 
\left( \frac{ T \Delta(T)}{ F } \right)^{\frac{1}{(\zeta_{eq} +1-\psi)}}
\label{letoite}
\end{eqnarray}
and the corresponding barrier scale
\begin{eqnarray}
B^*(F,T) \opsimeq_{F \to 0} \Delta(T)
\left( \frac{ T \Delta(T)}{ F } \right)^{\mu}
\label{betoite}
\end{eqnarray}
where $\mu$ is the exponent that usually appear in the creep formula
(see Eq. \ref{muusual} and associated comments)
\begin{eqnarray}
\mu \equiv \frac{\psi}{(\zeta_{eq} +1-\psi)}
\label{muetoite}
\end{eqnarray}
So the power-law exponent of Eq. \ref{alphaft2} vanishes for $F \to 0$ as
\begin{eqnarray}
 \alpha(F,T)= \frac{1}{B^*(F,T)}\opsimeq_{F \to 0}  \frac{1}{\Delta (T) }
 \left( \frac{F}{ T \Delta(T)} \right)^{\mu}
\label{alphafinal}
\end{eqnarray}

In the case of the Sinai model, one may actually use 
exactly the same arguments with the following changes :
the length $l$ is now along the direction of the motion,
the barrier exponent $\psi=1/2$ simply describes the Brownian
fluctuation of the random energy landscape, the prefactor $\Delta(T)$
has the simple $T$-dependence $\Delta(T)=C/T$ because the barriers
are purely energetic, and there is no roughness exponent $\zeta_{eq}=0$
since the model concerns a single particle.
For this special case, one obtains $\mu=1$
and the following expression for the power-law exponent
\begin{eqnarray}
 \alpha_{Sinai}(F,T) \propto T F
\label{alphasinai}
\end{eqnarray}
in agreement with the exact results mentioned at the beginning
of the introduction.

\subsection{ Final picture for the dynamics at small external force $F$  } 

In the case of the Sinai model, the saturation scales
$l^*(F,T)$ for the length and $B^*(F,T)$ for the barrier
are the only finite scales present in the problem.
The strong disorder renormalization procedure should then
be stopped at the scale $B^*(F,T)$, and one ends up with
an effective directed trap model. We refer to \cite{sinaibiasdirectedtraprg}
for a more detailed presentation of the
 quantitative relations that can be derived at large scales 
between the Sinai model in an external force
and the one-dimensional directed trap model.

For the present model concerning the directed polymer,
the polymer length $L$ introduces another length scale in the problem.
Of course, one is interested into large polymer length $L \gg 1$,
but since the length $l^*(F,T)$ diverges as $F \to 0$ (Eq. \ref{letoite}),
one needs to distinguish various regimes in terms of the polymer length $L$

\subsubsection{ Regime $1 \ll L \ll l^*(F,T)$ }

In the regime $1 \ll L \ll l^*(F,T)$, the relevant barriers for the motion
of the polymer as a whole corresponds to
the RG scale $\Gamma_L$ of Eq. \ref{gammaLstop},
which is well below the saturation scale
$B^*(F,T)$ introduced by the force $F$.
This means that the force $F$
 is then too small to impose a directed motion between two neighboring
quasi-equilibrated metastable configurations.
The motion will become effectively directed only on larger transversal
lengthscales separating many metastable configurations.

\subsubsection{ Regime  $L \sim l^*(F,T)$}

For a polymer of length of order $L \sim l^*(F,T)$,
 the motion of the polymer corresponds to an effectively directed
trap model. Each trap of trapping time $\tau_i$ corresponding to 
a transverse displacements of order $x_i= (l^*(F,T))^{\zeta} X_i$
(where $X_i$ is a random variable of order $O(1)$
with a finite averaged value $\overline{X_i}<\infty$).
After $n$ traps, the total displacement of the center of mass
 will be of order
\begin{eqnarray}
h_G(n) = x_1+x_2+...+x_n \opsimeq_{n \to \infty}
 n \left[ l^*(F,T) \right]^{\zeta} 
\label{XGtot}
\end{eqnarray}
whereas the total time needed to escape from these $n$ traps
will be of order 
\begin{eqnarray}
t(n) = \tau_1+\tau_2+...+\tau_n
\label{timetot}
\end{eqnarray}
As is well known in the field of L\'evy statistics \cite{jpbreview},
the sum of $n$ variables distributed with the broad power-law of Eq.
\ref{powerlaw2} grows linearly in $n$ only when the average value
$\overline{\tau}$ remains finite, i.e. for $\alpha>1$
\begin{eqnarray}
t(n)  \opsimeq_{n \to \infty} n \overline{\tau} \ \ {\rm if } \ \ \alpha(F,T)>1
\label{timetotmoyen}
\end{eqnarray}
However for $\alpha<1$, the average value of the trapping time diverges
$\overline{\tau}=+\infty$ and the sum grows more rapidly as
\begin{eqnarray}
t(n)  \opsimeq_{n \to \infty} n^{\frac{1}{\alpha}} \ \ {\rm if }
\ \  \alpha(F,T)<1
\label{timetotanomalous}
\end{eqnarray}
It is then clear that the asymptotic velocity is finite only for 
 $\alpha(F,T)>1$.
For $\alpha(F,T)<1$, the velocity vanishes $V=0$ and the center-of-mass
displacement grows only sub-linearly in time
\begin{eqnarray}
h_G(t)  \opsimeq_{t \to \infty} t^{\alpha(F,T)} \ \ {\rm if }
\ \  \alpha(F,T)<1
\label{xganomalous}
\end{eqnarray}

\subsubsection{ Regime $L \gg l^*(F,T)$  }

\label{interactingtrap}

Finally in the regime $L \gg l^*(F,T)$, 
the strong disorder RG procedure has to be stopped at 
the RG scale $B^*(F,T)$ of Eq. \ref{betoite}
corresponding to the length scale
$l^*(F,T)$ of Eq. \ref{letoite}.
The picture that emerges is then some kind of directed 'parallel' trap model,
where a number of order $L/l^*(F,T)$ segments of typical length 
$l^*(F,T)$ have to move in parallel
in a trap landscape where the statistics of trapping
times is given by the broad power-law of Eq. \ref{powerlaw2}
with exponent given in Eq. \ref{alphafinal}.
The main question is then 
to determine the form of the effective interaction
between these large segments from the knowledge of the microscopic 
interactions between monomers. 
This issue is discussed in detail in Appendix \ref{appendix}, 
where we first recall why the metric constraint is 
a safe choice for the microscopic interactions
(in contrast with a quadratic elastic energy that is known
to lead to an unphysical roughness exponent in some cases),
and where we then explain why the metric constraint is 
also an appropriate choice for the effective interactions
between consecutive segments of size $l^*(F,T)$.

From the two properties : 

(i)  each segment of size $l^*(F,T)$ would follow the sub-linear
 motion of Eq. \ref{xganomalous} if it were alone

(ii) the metric constraint between these segments
 can only delay (but not accelerate) each segment with respect to the motion
 it would have followed if it had been alone

we conclude that the center-of-mass motion will also
 follow the asymptotic sublinear behavior of Eq. \ref{xganomalous}
\begin{eqnarray}
h_G(t)  \opsimeq_{t \to \infty} t^{\alpha(F,T)} \ \ {\rm if }
\ \  \alpha(F,T)<1
\label{xganomalousbis}
\end{eqnarray}

 This is confirmed by the numerical simulations presented
 in the next section, where we moreover discuss 
the behavior of the interface width at the largest scales
 (i.e. at scales much larger than $l^*(F,T)$ for the initial microscopic model).

\section{ Numerical study of a directed polymer driven in a traps landscape }

\label{numerics}

In this section, we define and study
 an effective directed model within a landscape made of traps,
which is expected to be the appropriate coarse-grained model
of the true microscopic non-directed dynamics in the regime
$L \gg l^*(F,T)$ discussed above in section \ref{interactingtrap}.
We stress that a single monomer of this effective model
represents a quasi-equilibrated segment of length $l^*(F,T)$
of internal roughness $\zeta_{eq}=2/3$ of the true microscopic model
(see the previous sections for more explanations).

\subsection{ Definition of the effective directed model
 within a traps landscape }

We consider a directed polymer of length $L$
defined by the heights $(h_1,h_2,...,h_L)$ with cyclic boundary conditions $h_{L+1} \equiv h_1$
and with the metric constraint
\begin{eqnarray}
\vert h_{i+1}-h_i \vert = \pm 1
\label{chain}
\end{eqnarray}
The initial configuration is the 'flat' zig-zag configuration
\begin{eqnarray}
h_{2i} && =1 \nonumber  \\
h_{2i+1} && =0
\label{initial}
\end{eqnarray}
The traps landscape is defined as follows :
the random trapping times $\tau(i,h_i)$ 
are independent and drawn from the power-law distribution 
\begin{eqnarray}
q(\tau)
= \theta(\tau>1)  \frac{\alpha}{\tau^{1+\alpha}} 
 \label{qtau}
\end{eqnarray}

The dynamics is defined by a 
 Master Equation of the form of Eq. \ref{master}.
To respect the chain constraints of Eq. \ref{chain},
the monomer $(i)$ of the configuration ${\cal C} =(h_1,h_2,...,h_L)$
is 'movable' only if $h_{i-1}=h_i+1=h_{i+1}$, and the possible movement
is then $h_i \to h_i+2$.
The total rate out of the configuration ${\cal C}$ is thus given by
\begin{eqnarray}
W_{out}({\cal C} = \{h_1,h_2,...,h_L\}) && = 
 \sum_{i=1}^L W_i({\cal C}) \\
 W_i({\cal C} =\{h_1,h_2,...,h_L\}) && = \frac{\delta_{h_{i-1},h_i+1} 
\ \delta_{h_{i+1},h_i+1} }{\tau(i,h_i)}
\label{wouttrap}
\end{eqnarray}

The two main observables in the field of interface dynamics
are \cite{book_surfacegrowth}

(i) the height $h_G(t;L)$ of the center-of-mass of an interface
of length $L$ as a function of time $t$
\begin{eqnarray}
h_G(t;L) = \frac{1}{L} \sum_{i=1}^L h_i(t)
\label{hgt}
\end{eqnarray}

(ii) the width $w(t)$ of the interface of length $L$ as a function of time $t$
\begin{eqnarray}
w^2(t;L) = \frac{1}{L} \sum_{i=1}^L \left[h_i(t)-h_G(t) \right]^2
\label{w2t}
\end{eqnarray}

\subsection{ Numerical details  }

In our numerical study, we have used the
 'Bortz-Kalos-Lebowitz algorithm' \cite{algoBKL} which is a
faster-than-the-clock algorithm where each program iteration computes
the time and the site where the next movement occurs  \cite{werner}.
The idea is that from the knowledge of $W_{out}$,
the escape time $t_{esc}$ from configuration  ${\cal C}$
is a random variable drawn from the law
\begin{eqnarray}
P^{exit}_{\cal C} (t_{esc}) =
 W_{out}({\cal C}) e^{ -  W_{out}({\cal C}) t_{esc}} 
\label{pexit}
\end{eqnarray}
Then the monomer $(i)$ which is effectively moved $h_i \to h_i+2$
is drawn with the probability
\begin{eqnarray}
\pi(i)  = 
\frac{W_i({\cal C})}{W_{out}({\cal C})}
\label{pconfigjump}
\end{eqnarray}
In this framework, the total displacement $h_G$ of the center-of-mass 
is the natural variable,
and one computes for a given dynamics in a given disordered sample
the time $t(h_G)$ needed to reach a given displacement $h_G$
 of the center-of-mass.
Similarly, one computes the width $w(h_G)$ of the interface
 for a given displacement $h_G$ of the center-of-mass
starting from the flat initial condition of Eq. \ref{initial}. 
The numerical results presented below
correspond to measures at the values $h_G=2,4,....h_G^{max}$,
where the maximal displacement $h_G^{max}$ 
wished at the end of the simulation fixes the CPU time.

For the case $\alpha=0.5$ where the averaged trapping time diverges 
$\overline{\tau}=+\infty$ ( Eq \ref{qtau}),
we have made studies with $h_G^{max} \sim L$ and with $h_G^{max} \sim L^{1.5}$
for the following sizes and the corresponding number $n_s(L)$
of disordered samples :

(i) for $h_G^{max} \sim L$, we have studied sizes in the range 
$10^3 \leq L \leq 24.10^3$
with a statistics between $n_s(L=10^3) = 4.10^5$ and 
$n_s(L=24.10^3) = 1600$.

(ii) for $h_G^{max} \sim L^{1.5}$, we have studied sizes in the range 
$10^2 \leq L \leq 4.10^3$
with a statistics between $n_s(L=10^2) = 3.10^6$ and 
$n_s(L=4.10^3) = 500$.

For the case $\alpha=2$ where the averaged trapping time is finite 
$\overline{\tau} < +\infty $ ( Eq \ref{qtau}),
we have made studies with  $h_G^{max}  \sim L$
for sizes between  $10^2 \leq L \leq 10^4$
with a statistics between $n_s(L=10^2) = 4.10^5$ and 
$n_s(L=10^4) = 1600$. (Note that all other parameters being the same,
the CPU time turns out to be larger for the case $\mu=2$
than for the case $\mu=0.5$, because the number of 'movable monomers'
in the stationary regime is larger for the case $\mu=2$)

We give below results for single histories in a given disordered
sample, as well as disorder-averaged results
 denoted by an overbar.
In particular, we will be interested in
$\overline{ \ln t (h_G)}$ that represents the disorder-average
of the logarithm of the time needed to reach a displacement $h_G$
of the center-of-mass. 
For the width defined in Eq. \ref{w2t}, 
we moreover introduce the following notation
\begin{eqnarray}
w_{av}(t,L) \equiv \left( \overline{ w^2(t;L) } \right)^{1/2}
\label{wav}
\end{eqnarray}

\begin{figure}[htbp]
\includegraphics[height=6cm]{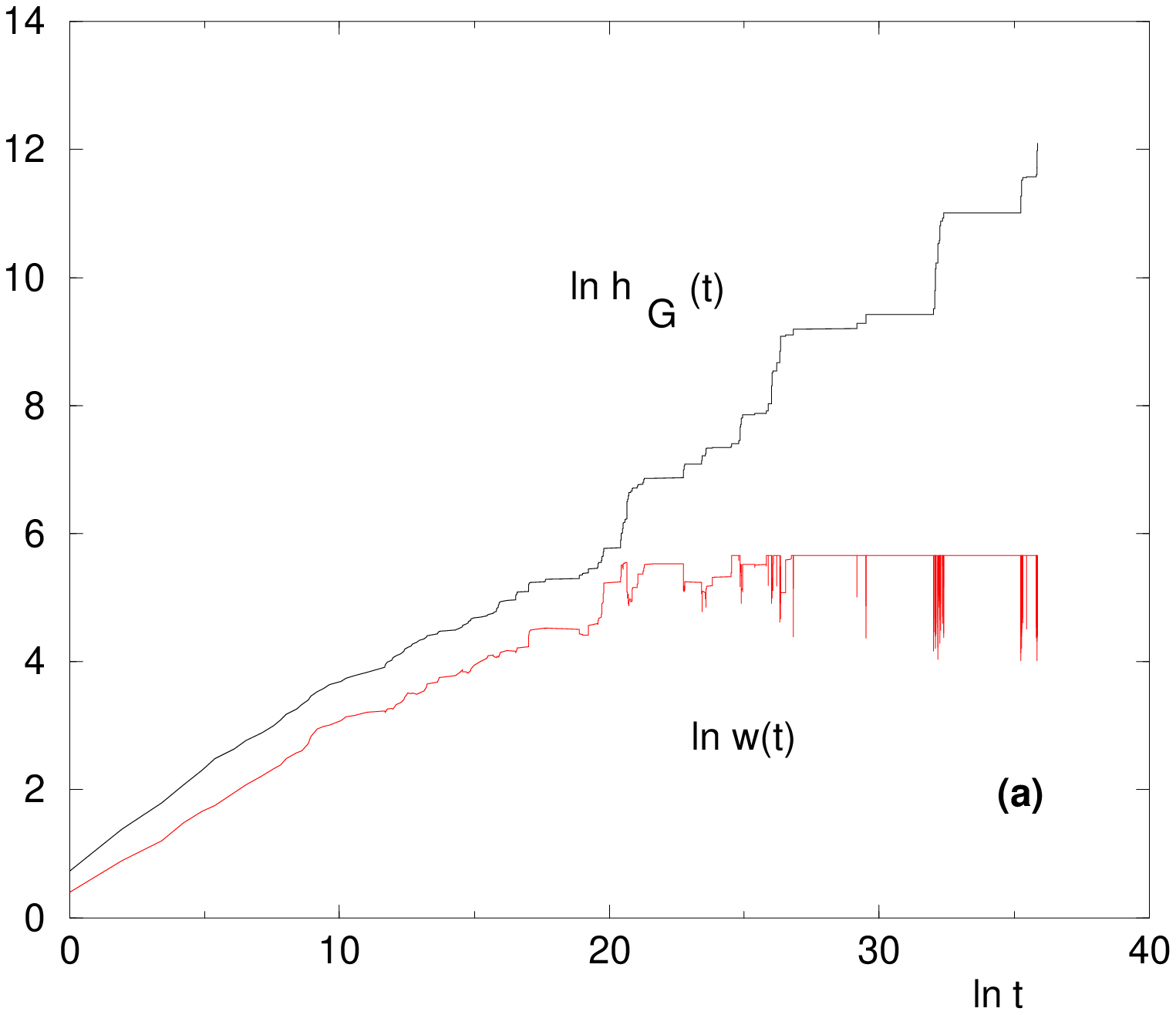}
\hspace{1cm}
\includegraphics[height=6cm]{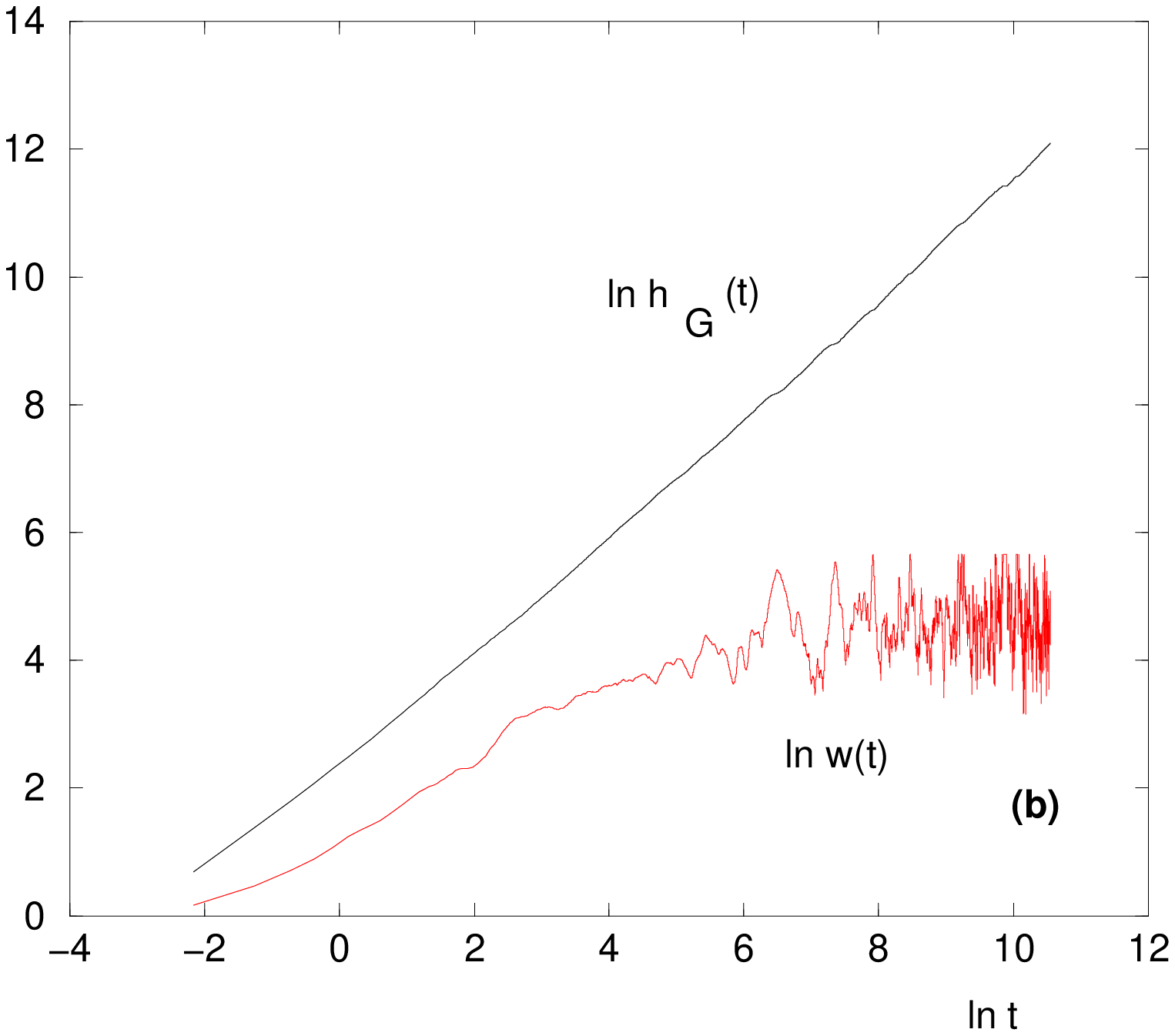}
\caption{ (Color online)
Typical dynamics of a single interface : log-log plot
 of the center-of-mass
height $h_G(t)$ and of the interface width $w(t)$ as a function of 
the time $t$ time
for a polymer of length $L=2000$
(a) for the case $\alpha=0.5$ where the averaged trapping time diverges 
$\overline{\tau}=+\infty$ ( Eq \ref{qtau}), the 
dynamics is very intermittent.
(b) for the case $\alpha=2$ where the averaged trapping time is finite 
$\overline{\tau}<+\infty$ ( Eq \ref{qtau}), the center-of-mass motion
is smooth. }
\label{singleruns}
\end{figure}

\subsection{ Typical dynamics of a single interface  }

We first consider a single history in a given disordered sample
 for a polymer of length $L=2000$.
 The dynamics of the center-of-mass position
 $h_G(t)$ (Eq. \ref{hgt}) and of the width $w(t)$ (Eq. \ref{w2t})
are shown on Fig. \ref{singleruns} (a)  for the case $\alpha=0.5$ 
 where the averaged trapping time diverges 
$\overline{\tau}=+\infty$ ( Eq \ref{qtau}),
and on Fig. \ref{singleruns} (b) for the case 
$\alpha=2$ where the averaged trapping time is finite 
$\overline{\tau}<+\infty$ ( Eq \ref{qtau}).
This comparison shows that the dynamics is very different
in the two cases.
For $\alpha=0.5$, the stationary regime is characterized 
by a very intermittent dynamics of the center-of-mass $h_G(t)$
and of the width $w(t)$, which remain pinned for long time intervals
separated by rapid avalanches.
On the contrary for $\alpha=2$, the motion of the center-of-mass
is smooth, and in the stationary regime, the width $w(t)$
fluctuates rapidly around its time-averaged value.

\begin{figure}[htbp]
\includegraphics[height=6cm]{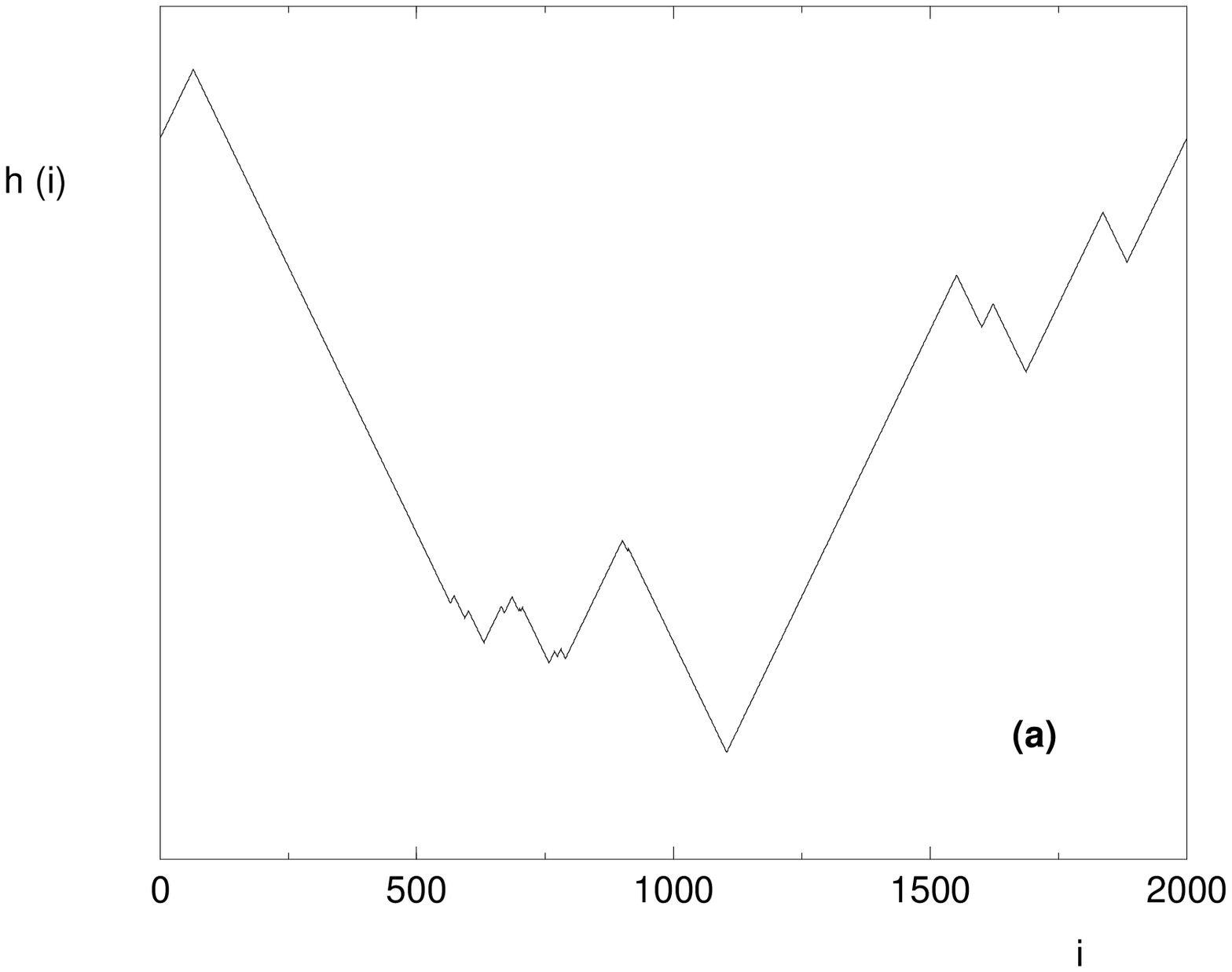}
\hspace{1cm}
\includegraphics[height=6cm]{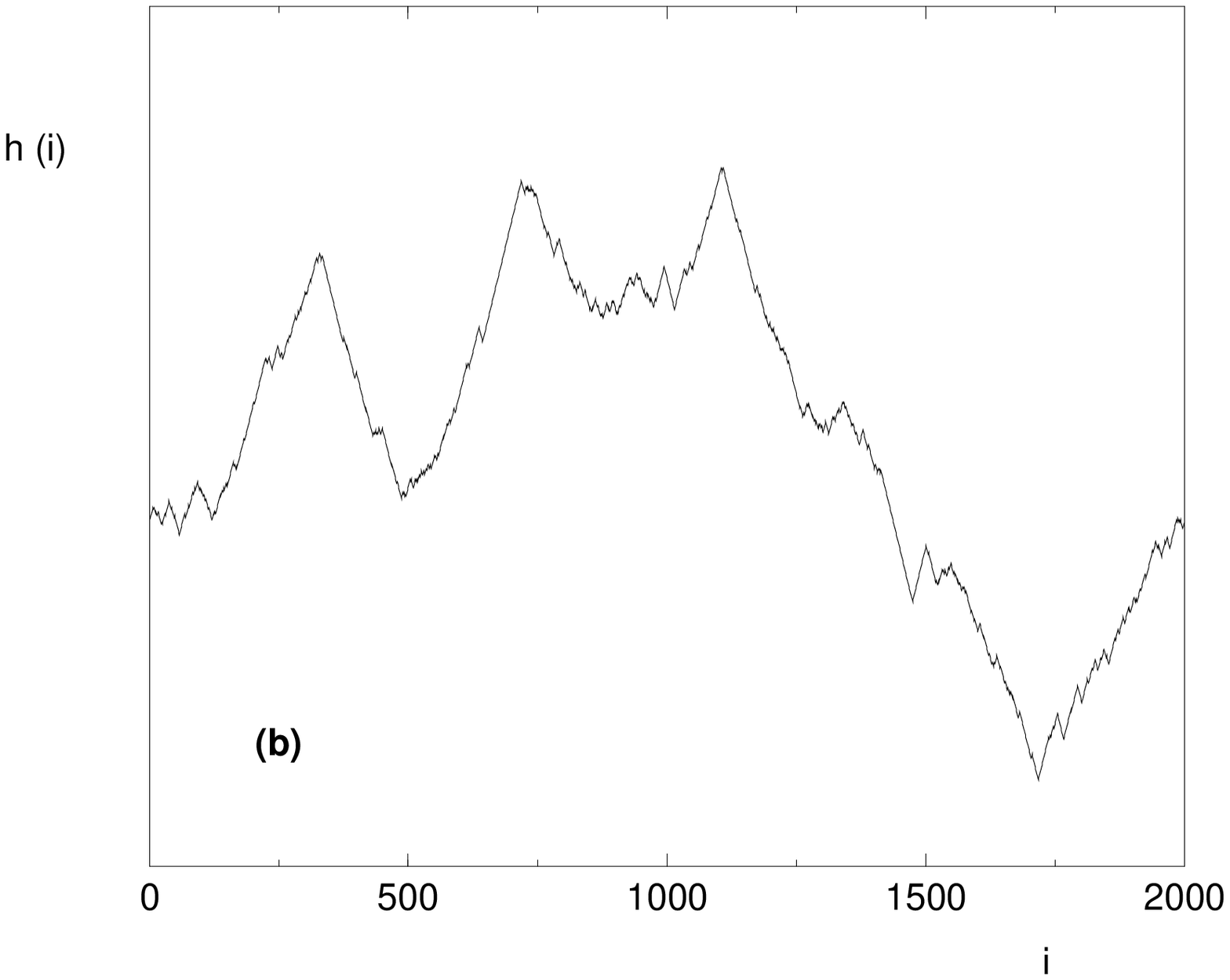}
\caption{ (Color online)
Typical configurations in the stationary regime
for a polymer of length $L=2000$
(a) for the case $\alpha=0.5$ where the averaged trapping time diverges 
$\overline{\tau}=+\infty$ ( Eq \ref{qtau}), very large regions of slope $1$
are present
(b) for the case $\alpha=2$ where the averaged trapping time is finite 
$\overline{\tau}<+\infty$ ( Eq \ref{qtau}) , there are more structures
on shorter scales. }
\label{config}
\end{figure}

We show on Fig. \ref{config} typical corresponding configurations
in the stationary regime.
For $\alpha=0.5$, one clearly see on Fig. \ref{config} (a) 
very large regions of slope $1$, whereas for $\alpha=2$,
the configuration of Fig. \ref{config} (b) presents more structures
on shorter scales.

\subsection{ Disorder-averaged behaviors for a given size $L$  }

\begin{figure}[htbp]
\includegraphics[height=6cm]{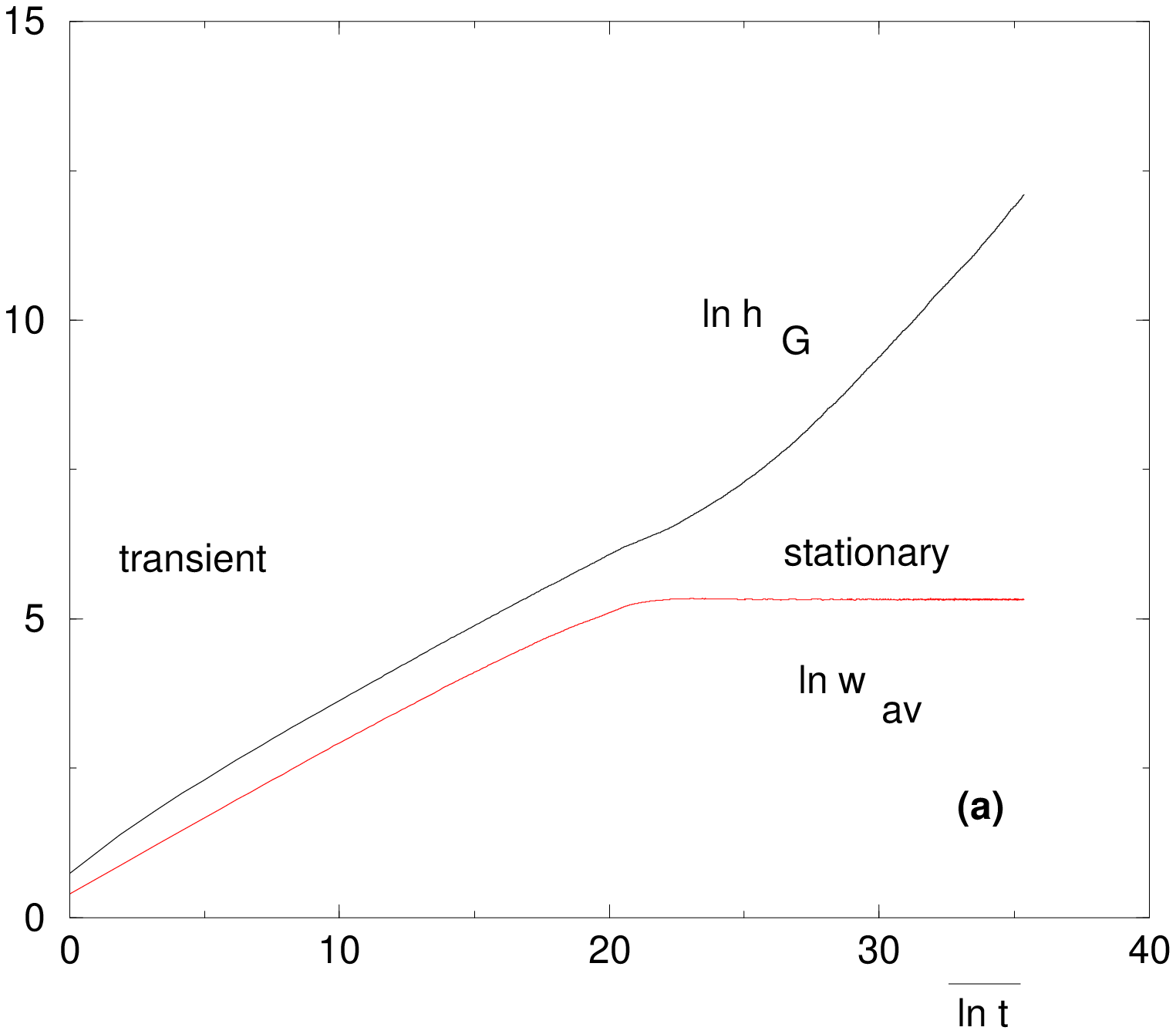}
\hspace{1cm}
 \includegraphics[height=6cm]{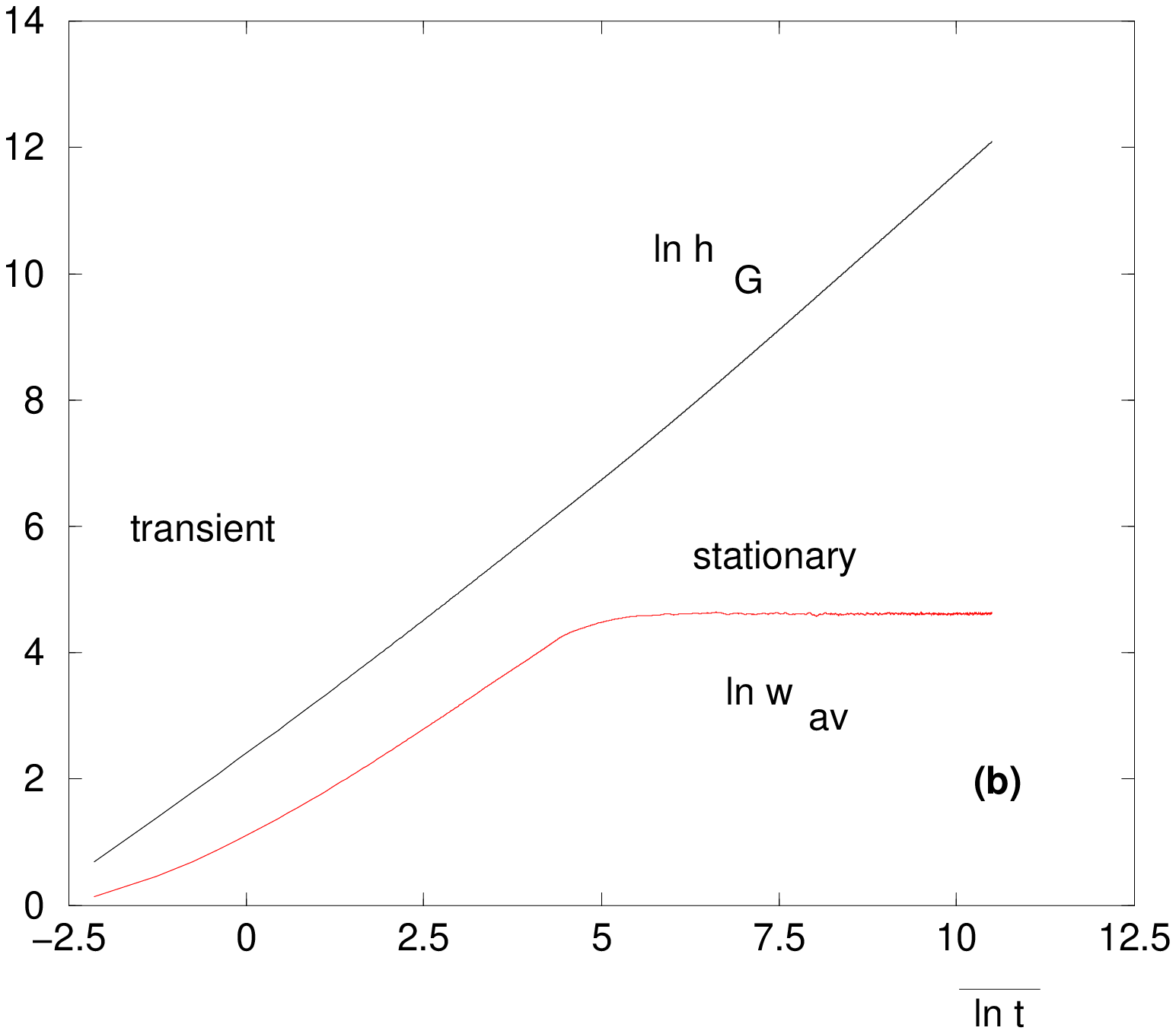}
\caption{ (Color online)
Disorder-averaged properties of the center-of-mass
height $h_G$ and of the interface width $w$ as a function of time
 (log-log plot) : data averaged over
$n_s=2000 $ disordered samples for a polymer of length $L=2000$
(a) for the case $\alpha=0.5$ where the averaged trapping time diverges 
$\overline{\tau}=+\infty$ ( Eq \ref{qtau}), the center-of-mass motion 
is sublinear with exponent $\beta \sim 0.25$ 
in the transient regime (Eq. \ref{slopebeta})
and with exponent $\alpha=0.5$ in the stationary regime (Eq. \ref{slope0.5})
(b) comparison with the 
case $\alpha=2$ where the averaged trapping time is finite 
$\overline{\tau}<+\infty$ ( Eq \ref{qtau}).
 }
\label{average2000}
\end{figure}

We have shown above on Fig \ref{singleruns}
 the dynamics of $h_G(t)$ and $w(t)$
for a single history of a polymer of size $L$.
We show on Fig. \ref{average2000} the same observables after
averaging over $n_s=2000$ disordered samples.

We note $t^*(L)$ the crossover time between the transient
regime where the disorder-averaged width $w_{av}(t,L)$
( Eq. \ref{wav}) grows and the stationary regime
where the disorder-averaged width saturates
 towards a time-independent value
\begin{eqnarray}
w_{av}(t,L) \opsimeq_{t \gg t^*(L)} w_{sat}(L)
\label{wavsat}
\end{eqnarray}

For $\alpha=0.5$, we find that in the stationary regime,
the plot of $\ln h_G$ as a function of $\overline{ \ln t (h_G)}$ 
(see Fig. \ref{average2000} a) corresponds to
a slope $0.5$
\begin{eqnarray}
\ln h_G \opsimeq_{t \gg t^*(L)} 0.5 \ln t +...= \alpha \ln t+...
\label{slope0.5}
\end{eqnarray}
in agreement with the sub-linear motion predicted in
 Eq. \ref{xganomalousbis}.
Of course for $\alpha=2$ 
we find that in the stationary regime,
the plot of $\ln h_G$ as a function of $\ln t$ 
(see Fig. \ref{average2000} b) corresponds to
a slope $1$ corresponding to the usual case of finite velocity
$h_G(t) \sim V t$.

For $\alpha=0.5$, we moreover measure that,  
in the initial transient regime, 
 the center-of-mass and the width both grows
with the same exponent $\beta$ in time
\begin{eqnarray}
\ln h_G(t,L) && \opsimeq_{t \ll t^*(L)}  \beta \ln t+... \nonumber \\
\ln w_{av}(t,L) && \opsimeq_{t \ll t^*(L)}  \beta \ln t+...
\label{slopebeta}
\end{eqnarray}
with a value of order
\begin{eqnarray}
\beta \sim 0.24
\label{mesurebeta}
\end{eqnarray}
that will be interpreted in section \ref{argument} below.

\subsection{ Roughness exponent in the stationary regime  }

\begin{figure}[htbp]
\includegraphics[height=6cm]{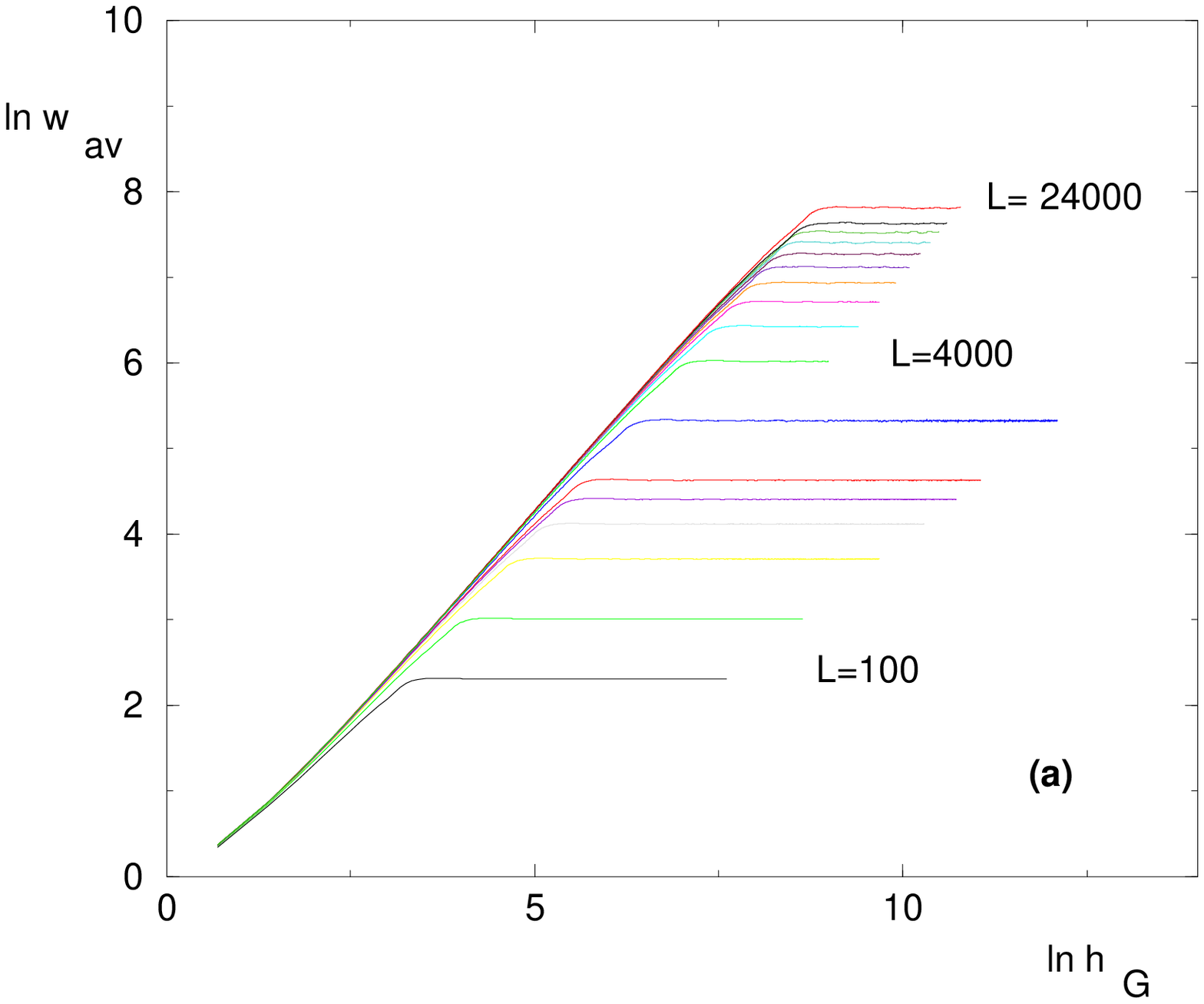}
\hspace{1cm}
 \includegraphics[height=6cm]{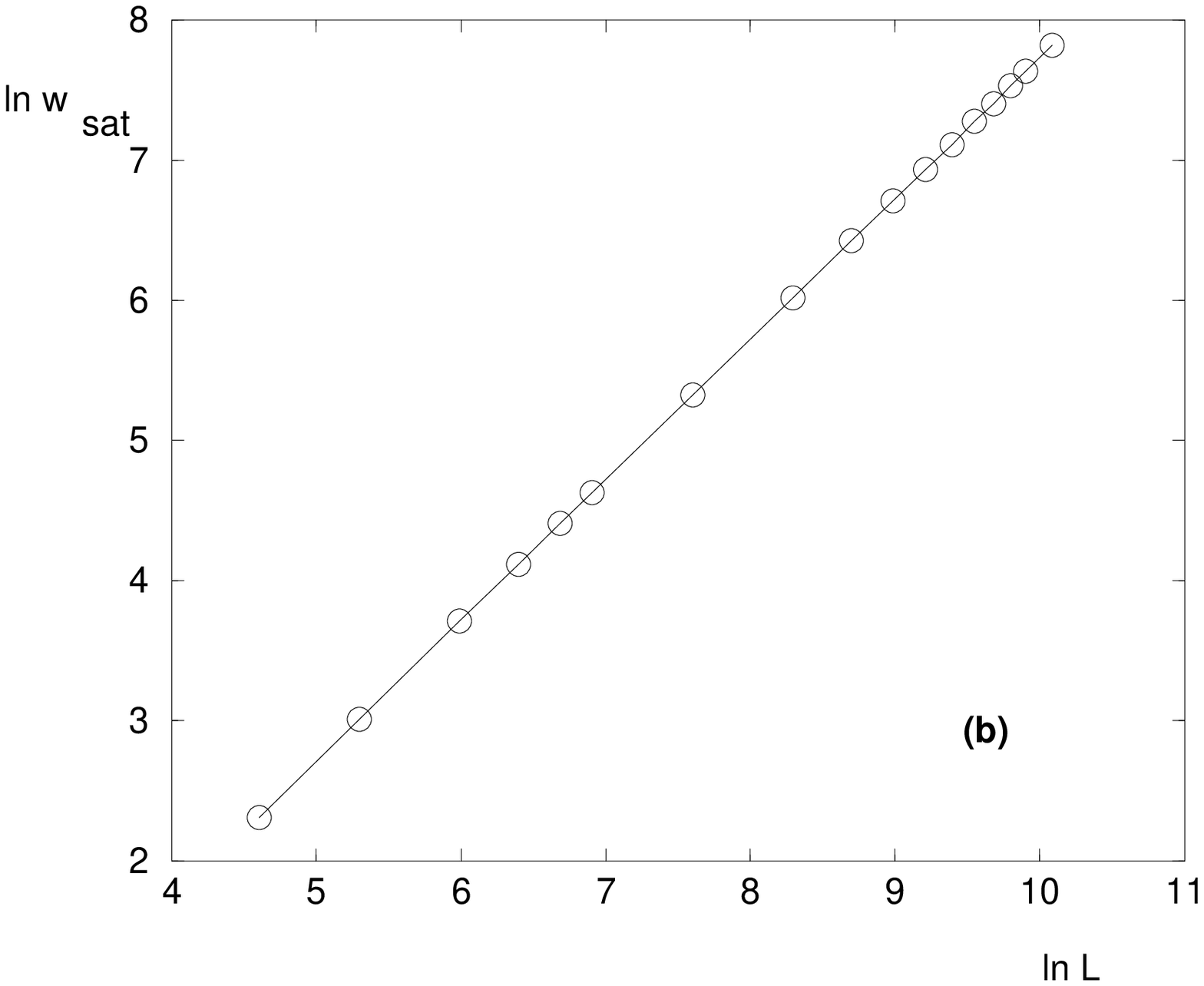}
\caption{ (Color online)
Dynamics of the width of the interface for $\alpha=0.5$ :
(a) $\ln w_{av}$
 as a function of $\ln h_G$ for various sizes $L$
(b) $\ln  w_{sat}(L) $ as a function of $\ln L$ :
the slope corresponds to the roughness exponent $\zeta=1$.
 }
\label{figroughnessmu0.5}
\end{figure}

We now consider the dynamics of the disorder-averaged width defined
in Eq. \ref{wav} as a function of the
center-of-mass displacement $h_G$.

For $\alpha=0.5$, we show on Fig. \ref{figroughnessmu0.5} (a)
 the dynamics of the width $w_{av}$ for various polymer sizes $L$. 
We show on Fig. \ref{figroughnessmu0.5} (b) the log-log plot
of the saturation value $w_{sat}(L)$ defined in Eq. \ref{wavsat}
as a function of $L$ : we find a slope of order $1$
\begin{eqnarray}
\ln w_{sat} (L) \simeq \ln L +... 
\label{slope1}
\end{eqnarray}
Our conclusion is thus that for $\alpha<1$,
the roughness exponent is $\zeta =1$,
as already suggested by the shape of typical configurations
in the stationary regime (see Fig. \ref{config} a )

\begin{figure}[htbp]
\includegraphics[height=6cm]{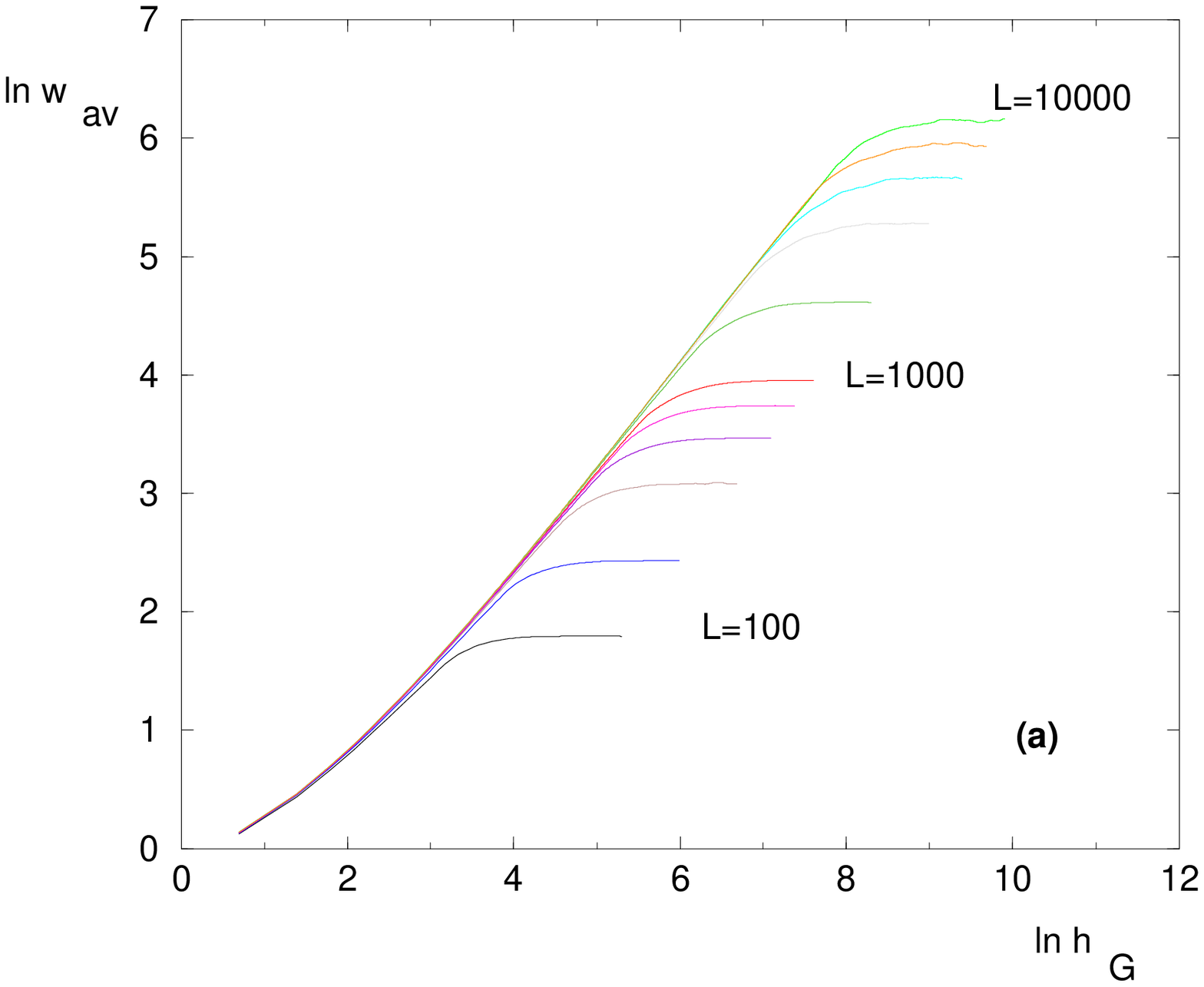}
\hspace{1cm}
 \includegraphics[height=6cm]{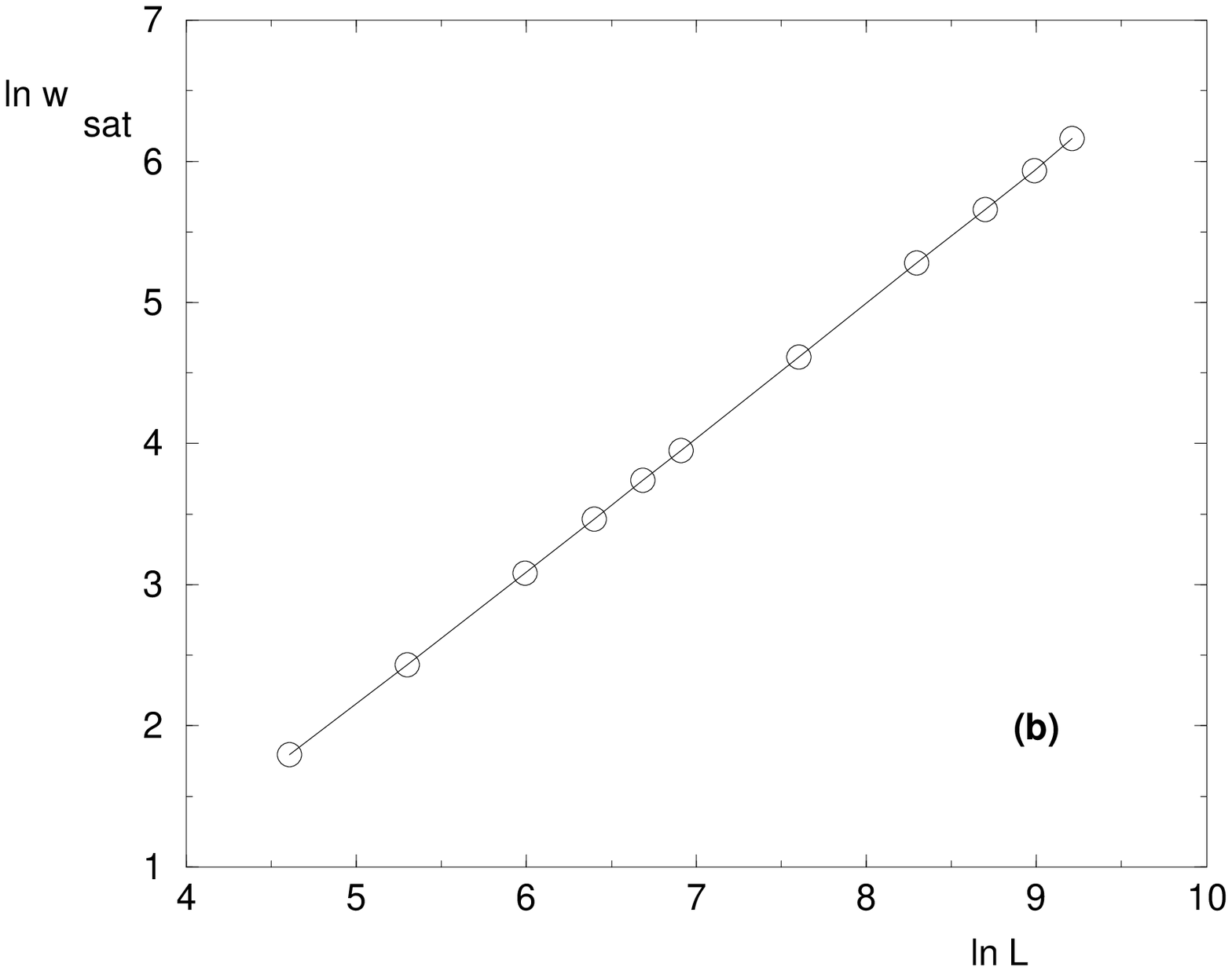}
\caption{ (Color online)
Dynamics of the width of the interface for $\alpha=2$
(a) $\ln w_{av}$
 as a function of $\ln h_G$ for various sizes $L$
(b) $\ln  w_{sat}(L) $ as a function of $\ln L$ :
the slope is of order $0.95$. }
\label{figroughnessmu2}
\end{figure}

For comparison, we show on Fig. \ref{figroughnessmu2}
 the same observables for the case $\alpha=2$.
In particular, the log-log plot of 
 Fig. \ref{figroughnessmu2} (b) 
corresponds to a slope of order 
$\ln w_{sat}(L) \sim 0.95 \ln L $
 for the width in the stationary regime.
This high value can be understood from 
the typical configuration
in the stationary regime shown on Fig. \ref{config} b.
This suggests that even in the finite velocity phase $\alpha>1$
where the averaged trapping time is finite (Eq \ref{qtau}),
 it is very difficult for the different regions of the polymer to remain
synchronized during the dynamics, 
so that the the polymer tends to be stretched on large distances with 
a slope close to the maximal slope $1$ allowed in the presence of the metric
constraint.

\subsection{ Finite-size scaling forms in $t$ and $L$ in the case $\alpha<1$  } 

\begin{figure}[htbp]
\includegraphics[height=6cm]{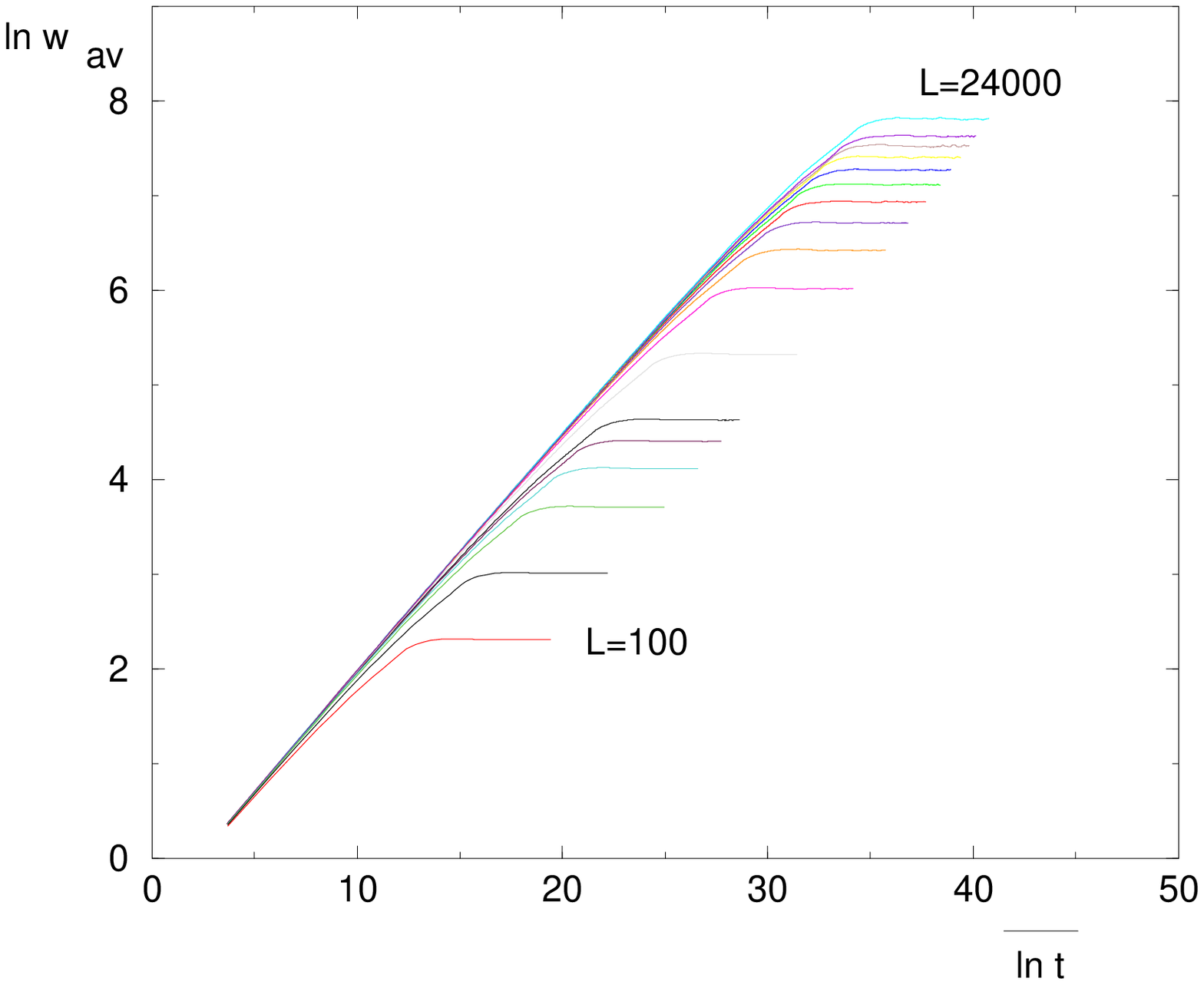}
\hspace{1cm}
 \includegraphics[height=6cm]{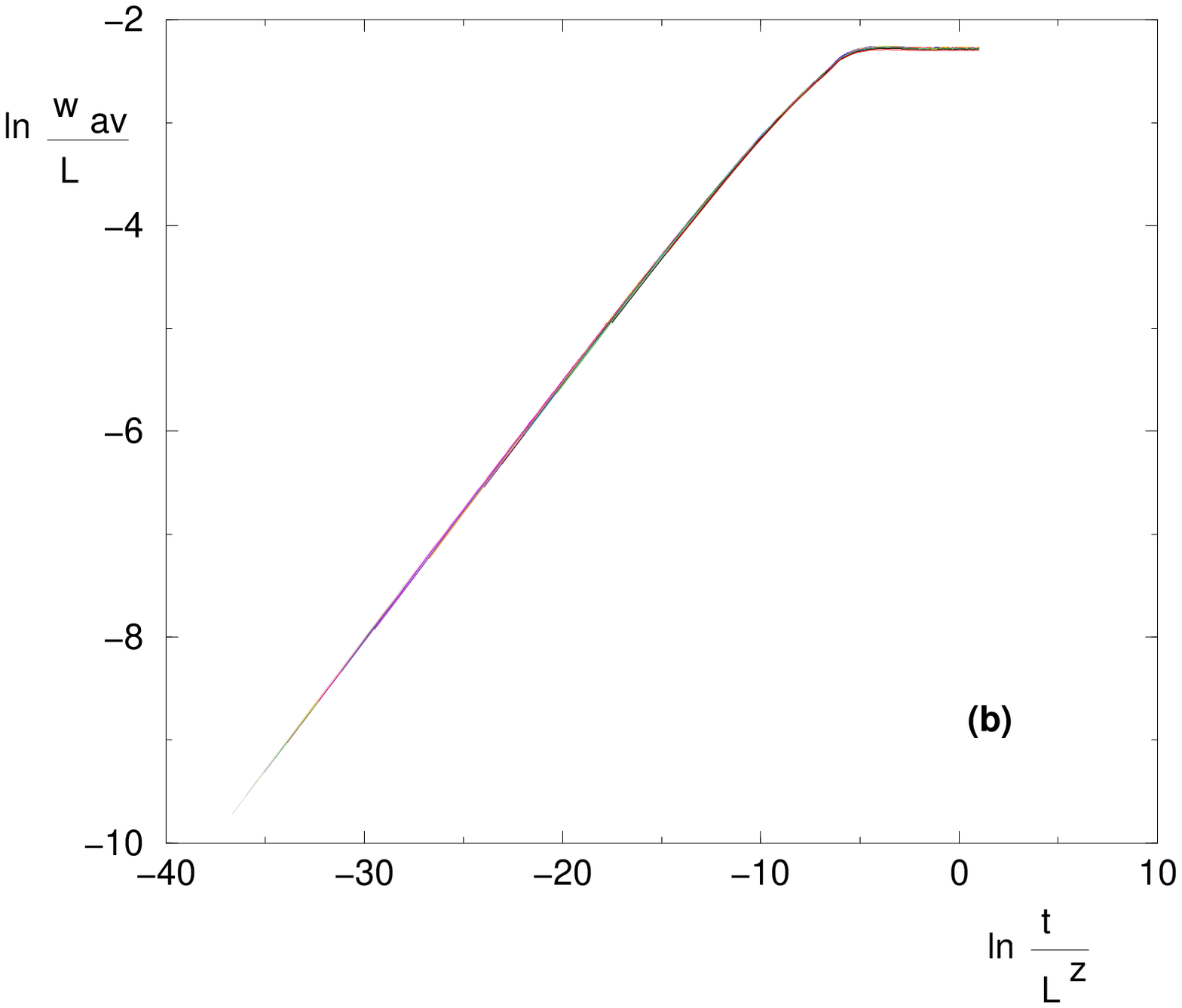}
\caption{ (Color online)
Finite-size scaling analysis in $t$ and $L$
 the width of the interface in the case $\alpha=0.5$ :
(a) $\ln w_{av}$
 as a function of $\ln t$ for various sizes $L$
(b) Data collapse of the same data using
the variables $\ln \left( w_{sat}(L)/L \right) $ as a function
 of $\ln \left( t/L^z \right) $ (see Eq. \ref{scalingwav}) with $z=4$. }
\label{fsswidth}
\end{figure}

In the field of interface dynamics, it is usual to summarize
the crossover between the transient dynamics and the stationary
dynamics by the following 
scaling form for the width \cite{book_surfacegrowth}
\begin{eqnarray}
 w_{av} (t;L) \simeq L^{\zeta} \phi \left( \frac{t}{L^z} \right) 
\label{scalingwav}
\end{eqnarray}
where $\zeta$ represents the roughness in the stationary regime,
where $z$ is the dynamical exponent of the crossover time $t^*(L) \sim L^z$
and where the scaling function $\phi(x)$ has the following behaviors :

(i) it converges towards a constant at infinity $\phi(x \to \infty)
 \sim const$ 
to recover the saturation value (Eq. \ref{wavsat}) at large time ;

(ii) it behaves as a power-law at small argument : $\phi(x \to 0)
 \sim x^{\beta}$ where the exponent $\beta$ governs
the growth in time during the initial transient behavior
\begin{eqnarray}
 w_{av} (t;L) \opsimeq_{t \ll  L^z }
L^{\zeta- \beta z} t^{\beta} \sim t^{\beta}
\label{transientwav}
\end{eqnarray}
where the scaling relation $\zeta=\beta z$ between exponents is expected
to obtain a $L$-independent behavior in the transcient regime \cite{book_surfacegrowth}.

We show on Fig. \ref{fsswidth} that a good data collapse
can be achieved with the values $\zeta=1$ and
\begin{eqnarray}
z(\alpha=0.5) \simeq 4
\label{znume}
\end{eqnarray}
The corresponding exponent of the scaling function $\phi(x)$ at small argument
is measured to be of order
\begin{eqnarray}
\beta(\alpha=0.5) \simeq 0.24
\label{betanume}
\end{eqnarray}
in agreement with the previous estimate of Eq. \ref{mesurebeta}.

Equivalently, as a function of the center-of-mass position $h_G$,
the width satisfies the finite-size scaling form 
 \begin{eqnarray}
 w_{av} (h_G;L) \simeq L \psi \left( \frac{h_G}{L} \right) 
\label{scalingwavhg}
\end{eqnarray}
where the scaling function $\psi(x)$ has for asymptotic behaviors
$\psi(x \to \infty) = const$ and $\psi(x \to 0) \propto x$,
so that in the transient regime, one has the simple proportionality
 \begin{eqnarray}
 w_{av} (h_G;L) \opsimeq_{ h_G \ll L} (const) \ \ h_G 
\label{scalingwavhginitial}
\end{eqnarray}

The compatibility between the finite-size scaling forms of Eq. \ref{scalingwav}
and of Eq. \ref{scalingwavhg} yields the following finite-size scaling form
for the center-of-mass position
 \begin{eqnarray}
 h_G (t;L) \simeq L H \left( \frac{t}{L^z} \right) 
\label{scalinghg}
\end{eqnarray}
where $H(x)$ is a scaling function (data not shown). 

\subsection{ Simple arguments to determine the exponents for $0<\alpha<1$  } 

\label{argument}

We expect that the roughness $\zeta=1$ found above numerically
for the special case $\alpha=1/2$ remains the same
in the whole phase $0<\alpha<1$ where strong non-self-averaging effects
of the trapping times (Eq \ref{qtau}) occur
 \begin{eqnarray}
\zeta(0<\alpha<1) =1
\label{zetavalue}
\end{eqnarray}

We may now use a simple argument to determine
 the other exponents as a function of
$\alpha$ in the interval $0<\alpha<1$.
For $h_G$ large, the time $t(h_G,L)$
 needed to make a center-of-mass displacement of order $h_G $
for a polymer of length $L$ may be estimated from the maximal
$\tau_{max}(N)$ trapping time among $N=(L h_G)$ independent variables drawn
with the probability distribution of Eq \ref{qtau}
 \begin{eqnarray}
 \tau_{max}(N) \sim  N^{1/\alpha} \sim (L h_G)^{1/\alpha}
\label{taumax}
\end{eqnarray}
From this maximal trapping time, the minimal rate transition rate 
encountered is from Eq. \ref{wouttrap}
 \begin{eqnarray}
W_{out}^{min} \sim \frac{1}{ \tau_{max}(L h_G)} \sim \frac{1}{ (L h_G)^{1/\alpha}}
\label{wmin}
\end{eqnarray}
Assuming that the time $t$ scales as the inverse of this minimal rate
$t \sim 1/W_{out}^{min} \sim  \times (L h_G)^{1/\alpha}$, we obtain 
by inversion the following scaling in the stationary regime
 \begin{eqnarray}
 h_G (t;L) \opsimeq_{t \gg t^*(L)}  \frac{t^{\alpha} }{ L}
\label{hgstatio}
\end{eqnarray}
The scaling function $H(x)$ of Eq. \ref{scalinghg} thus grows asymptotically
as the power-law $H(x \to \infty ) \sim x^{\alpha}$
and the dynamical exponent $z$ describing the scaling of the crossover
time $t^*(L) \sim L^z$ reads
 \begin{eqnarray}
z(\alpha) = \frac{2}{\alpha}
\label{zvalue}
\end{eqnarray}
For $\alpha=1/2$ used in our simulations, this corresponds to 
$z =4$ in agreement with the value measured above (Eqs \ref{znume}).

To determine the exponent $\beta$ of the initial transient regime,
the above argument has to be changed as follows.
We expect that in the transient regime,
an extensive number $(\rho L)$ of monomers
 (with a finite density $\rho$) are still
pinned in their initial positions. Equivalently, there is a number
$\rho L$ of segments of typical length $l_{\rho} \sim L/(\rho L) \sim 1/\rho $,
that have made a move of typical amplitude $h_i \sim l$,
because the dynamics is building the roughness $\zeta=1$ at these
small scales.
As a consequence, the center-of-mass displacement 
is of order $h_G \sim \rho l_{\rho}^2 \sim  l_{\rho}$.
Since all these segments are independent, we now have to evaluate
the time $t_{l_{\rho}}$ for a single segment of size $l_{\rho}$ to move
over a distance of order $l_{\rho}$, i.e. the time $t_{l_{\rho}}$ to overcome
 $l_{\rho}^2$ trapping times.
The maximal trapping time among these scales as  (see Eq. \ref{taumax})
  \begin{eqnarray}
 \tau_{max}( N_{\rho}= l_{\rho}^2 )  \sim  N^{1/\alpha}
 \sim l_{\rho}^{2/\alpha}
\label{taumaxbis}
\end{eqnarray}
The time $t_{l_{\rho}}$ can be thus estimated as
in Eq. \ref{wmin} : $t_{l_{\rho}}
 \sim 1/W_{out}^{min} \sim  \times l_{\rho}^{2/\alpha}$.
Using $h_G \sim l_{\rho}$ derived above, we obtain by inversion
the transient behavior 
 \begin{eqnarray}
 h_G (t;L) \opsimeq_{t \ll t^*(L)} t^{\frac{\alpha}{2}} 
\label{hgtransient}
\end{eqnarray}
The scaling function $H(x)$ of Eq. \ref{scalinghg} thus grows at the origin
as the power-law $H(x \to 0 ) \sim x^{\alpha/2}$
i.e. the transient exponent $\beta$ reads
 \begin{eqnarray}
\beta(\alpha) = \frac{\alpha}{2}
\label{betavalue}
\end{eqnarray}
For $\alpha=0.5$, 
this is in agreement with the value $\beta \sim 0.24$ measured 
 above (see Eqs \ref{mesurebeta} and \ref{betanume}).
Finally, with the values of Eqs  \ref{zetavalue}, \ref{zvalue} and 
 \ref{betavalue},
the transient behavior of the width of Eq. \ref{transientwav}
simplifies into the same scaling behavior as the center-of-mass
displacement $h_G$ (Eq. \ref{hgtransient})
\begin{eqnarray}
 w_{av} (t;L) \opsimeq_{t \ll t^*(L)} t^{\frac{\alpha}{2}} 
\label{transientwavfinal}
\end{eqnarray}
as expected from Eq. \ref{scalingwavhginitial}.

In conclusion, the dynamics in the zero-velocity phase $0<\alpha<1$
can be understood in terms of
simple arguments based on the statistics of trapping times.
The critical exponents that are computed from these arguments
are in agreement with our numerical simulations
both in the stationary regime and the transient regime.

\section{ Comparison with roughness exponents measured in previous works } 

\label{previous}

For a polymer with the metric constraint at finite
temperature driven by a small force, we may summarize 
 the behavior of the roughness discussed in the above sections as follows :

(i) at scales smaller than $l^*(T,F)$, the force is too small to really
change the energy landscape seen by the polymer at $F=0$, and thus
the roughness is expected to be governed by the equilibrium value
 $\zeta_{eq}=2/3$.

(ii) at scales larger than $l^*(T,F)$, we have found in our numerical study
of the previous section that, at small force where the exponent
of the trapping time power law (Eqs \ref{powerlaw2} and \ref{alphaft2})
is in the interval $0<\alpha(F,T)<1$, the roughness becomes $\zeta=1$.

In this section, we compare this scenario with previous works.

\subsection{ Relation with other works finding a roughness exponent $\zeta=1$ 
in the presence of a metric constraint }

We are aware of two works where a roughness exponent
 $\zeta=1$ has been measured for the dynamics
in the presence of a metric constraint:

(i) Sneppen has found a roughness exponent $\zeta=1$ for the following
 self-organized interface dynamics called 'model A' in \cite{Sneppen} :
 at each iteration, the site with the smallest pinning force among the sites that are allowed
 to move without breaking the metric constraint, moves forward by one unit and
 a new pinning force is drawn for this site.
 
 Self-organized-criticality models are usually not directly related to 
finite-temperature dynamics.
Here for instance, the important differences 
 with the effective directed traps model
 defined by the master equation of Eq. \ref{master}
with Eq. \ref{wouttrap} are the following : 
 in Sneppen's model, there is no real time,
 but only a number of iterations that corresponds to the total displacement $h_G$ of the center-of-mass;
and at each iteration, it is the always the
 least pinned movable site that moves, whereas in the master equation
 of Eq \ref{master}, the time and the site of the next moves are drawn with Eqs \ref{pexit} and \ref{pconfigjump}).
Nevertheless, there is clearly a close relationship between the two models,
in particular in the mechanism 
that generates a roughness exponent $\zeta=1$
in the stationary state 
 (see the typical configurations on Fig. \ref{config} (a) of
the present paper and Fig. 1a of Ref. \cite{Sneppen}).

(ii)  Tang and Leschhorn have found in their analysis 
of the dynamics slightly above the directed percolation depinning threshold  
\cite{TL_DPpinning,Leschhorn_anisotropic} 
 that the moving interface is not self-affine, 
but is a mixture of two kinds of behaviors :
there exists finite pinned segments that have for roughness 
the exponent of the depinning transition $\zeta_{dep}=\zeta_{DP} \sim 0.63$ 
whereas segments with a slope of order $1$ are found between them.

Since the region discussed by Tang and Leschhorn is 
on the horizontal axis of Fig. 2
slightly above the depinning critical point, it is very far from the region near the vertical axis of Fig. 2 discussed in the present paper.
However, the two pictures that emerge have nevertheless
some similarities : in our case, 
the local roughness is determined by the equilibrium
exponent $\zeta_{eq}=2/3$ up to the length $l^*$, in their case the local roughness is the depinning roughness $\zeta_{dep}=\zeta_{DP} \sim 0.63$,
but in both cases the roughness becomes equal to 1 at largest scales,
 i.e. the maximal roughness which is physically acceptable.
This seems to indicate that in both cases,
 it is very difficult for the different regions of the polymer to remain
synchronized during the dynamics, 
so that the maximal slope 1 allowed by the metric constraint 
is actually reached
to maintain the full polymer together.

\subsection{ Roughness exponents measured in other works in the presence of 
an elastic energy  } 

Most of the numerical studies on the dynamics at finite temperature
have been done for a Langevin dynamics in the presence of an elastic quadratic 
energy between monomers : 

(i) the equilibrium roughness exponent
 $\zeta_{eq}=2/3$  has been measured
during the driven dynamics 
in \cite{kaper}, and in \cite{Kol_tworegimes} when the temperature remains 
larger that some disorder-dependent threshold. 
In our opinion, this means that in these two simulations,
the length $L$ of the polymer was smaller or of the order
of the length $l^*(F,T)$.

(ii) at lower temperature in \cite{Kol_tworegimes}, the authors have measured
an effective roughness exponent of order $\zeta \simeq 0.9$.
In the study \cite{Kol_below} concerning the limit $T \to 0$, 
the authors have found a crossover between
the equilibrium value $\zeta_{eq}=2/3$ at small scales and 
the critical depinning value $\zeta_{dep} \sim 1.26$ at larger scales,
and have related this crossover to Functional RG calculations
\cite{chauve}.
Since any roughness exponent $\zeta>1$ is unphysical 
(see \cite{LT_roughnessunphysical,rosso_origin} 
and the discussion in Appendix A),
our opinion is that it would be very 
interesting if the simulations of \cite{Kol_below}
based on a quadratic elastic energy
 were done with the metric constraint instead,
to see what physically acceptable roughness exponent $\zeta \leq 1$ 
would actually emerge at large scales.

\section{Conclusions and perspectives}

\label{conclusion}

In this paper, we have considered the dynamics of the directed polymer
in a two-dimensional random potential in the regime where the temperature $T$
is finite and the external force $F$ is small.
We have explained how the ``infinite disorder fixed point'' that 
describes the dynamics for $F=0$ becomes a ``strong disorder fixed point''
for small $F$ with an exponential distribution
of renormalized barriers. Since the corresponding 
distribution of trapping times then only decays as a power-law
$P(\tau) \sim 1/\tau^{1+\alpha}$, 
where the exponent $\alpha(F,T)$ vanishes as $\alpha(F,T) \propto
F^{\mu}$ as $F \to 0$, we have concluded
that the motion is only sub-linearly in time 
$h_G(t) \sim  t^{\alpha(F,T)}$ in the region $\alpha(F,T)<1$,
i.e. that the asymptotic velocity vanishes $V=0$, in contrast with 
the usual creep scenario where the velocity is finite as soon as
 $(T>0,F>0)$.
All along the paper, we have discussed the similarities 
 with the Sinai model with bias,
 where an analogous zero-velocity phase has been established
long ago by rigorous methods 
\cite{kesten,derrida_pom,feigelman_vin,jpb_annphys}
and where the asymptotic exactness of the strong disorder renormalization
 has been demonstrated explicitly
by a direct comparison with the available rigorous results 
 \cite{sinairg,sinaibiasdirectedtraprg,review}.
We have then checked the presence of the predicted zero-velocity phase
by numerical simulations of a directed polymer
with a metric constraint driven in a traps landscape. 
We have moreover obtained that the roughness, which is governed by
 the equilibrium exponent $\zeta_{eq}=2/3$
up to the large scale $l^*$, is equal to $\zeta=1$ at the largest scales.

An important issue is of course whether such a zero-velocity phase also
exists for interfaces of higher dimensionalities $d>1$ driven in
a random medium of dimension $(d+1)$, and more generally 
for other classes of driven extended systems.
Since ``collective transport in random media
is an impossibly broad subject'' \cite{dsfreview},
a general answer clearly goes beyond the present work.
However we think that the essential property
needed to have a zero-velocity phase at small force
is the presence of a positive barrier exponent $\psi>0$
 for the dynamics at $F=0$, and this should be the case
for a broad class of disordered systems in finite dimensions.
 Then the dynamics for $F=0$ will be logarithmically slow
and should correspond to some ``infinite disorder fixed point''
for the renormalized barriers, that transforms into a
 ``strong disorder fixed point'' at small $F$, with an exponentially
distribution of renormalized barriers.
Another argument in favor of this general scenario
is the extremal statistics argument on the barriers \cite{rammal,Vin_Mar}
that also lead to an exponential tail for the probability
distribution of large barriers,
and thus to a power-law decay for the trapping time distribution. 

\section*{ Acknowledgements }

 It is a pleasure to thank J.P. Bouchaud for sending us a copy of his
 conference proceeding \cite{jpb_cargese}.

\appendix

\section{ Type of interactions : metric constraint or elastic energy }

\label{appendix}

In the text, the interaction used to maintain the continuity
of the interface is a metric constraint, both for the microscopic
model and for the effective directed trap model described in section
\ref{interactingtrap}.
In the present appendix, we explain this choice.

\subsection{ Subtleties in the choice of microscopic interactions} 

Let us first recall the two types of {\it microscopic interactions}
 that are usually considered in
the literature for a
 polymer described by the heights $\{h_i  \}$ of monomers :

\subsubsection{ Metric constraint between the heights of two neighboring monomers :  $\vert h_{i+1} -h_i \vert \leq 1$. }

This metric constraint is for 
instance very much used in numerical studies using transfer
matrix methods (see the review \cite{Hal_Zha}).
It can can be moreover justified in various microscopic models,
in particular when the directed polymer represents an interface in a two-dimensional
disordered ferromagnet well below the critical temperature $T_c$,
which was the original motivation to introduce the directed polymer model \cite{Hus_Hen}.

\subsubsection{ Elastic energy of the form :  $E_{el} (h_{i+1} -h_i) \sim (h_{i+1} -h_i)^n$, with $n=2$ usually. }

It is of course very common in physics to replace a hard constraint
by a soft constraint. In the present case,
the use of an elastic quadratic energy usually comes from 
a small-gradient expansion, an hypothesis which has
 to be consistent with the results obtained for the roughness.
Indeed, the roughness exponent $\zeta$ has to satisfy the bound $\zeta \leq 1$
for at least two reasons \cite{LT_roughnessunphysical,rosso_origin} :
 first, if one obtains $\zeta>1$,
the gradient is not small, and thus there is an inconsistency
with the small gradient expansion used to obtain the elastic quadratic energy ;
second, if one obtains $\zeta>1$, the elastic energy diverges
 in the thermodynamic limit $L \to \infty$.

\subsubsection{Discussion} 

The metric constraint can be seen as the limit $n \to \infty$ of the power $n$
of the elastic energy. As a consequence, 
there is room for various universality classes
between $n=2$ and $n=\infty$.
 So one should not assume a priori that all these interactions are equivalent,
but test for each case of interest if they lead to the same results or not.

For the directed polymer at equilibrium at any temperature $T$
 and no external force
in dimension $1+1$, the same equilibrium roughness 
exponent $\zeta_{eq} =2/3$ arises if one considers the metric constraint
or the elastic energy with $n=2$ (see the review\cite{Hal_Zha}).
However, at the zero-temperature depinning transition (see Fig. 2), 
some subtleties arise (see the detailed discussion in \cite{rosso_origin}) :
it turns out that the model defined in terms of an elastic quadratic energy 
(case $n=2$)
gives a roughness exponent $\zeta_{n=2} \sim 1.25$ which is unphysical
because it is strictly greater than one $\zeta_{n=2} >1$
\cite{LT_roughnessunphysical,rosso_origin}. 
This is in contrast with the model
defined either with the metric constraint or 
with an elastic energy of at least quartic order ($n \geq 4$)  
 where the roughness exponent is physical $\zeta_{metric} \leq 1$
and actually takes a value $\zeta_{metric} \sim 0.63 \sim \zeta_{n=4}$ 
\cite{rosso_origin}
that can be understood in terms of pinning by Directed Percolation clusters
 \cite{TL_DPpinning,buldyrev}.

In conclusion, since the microscopic model with an elastic
 quadratic energy (case $n=2$) is known to lead to 
unphysical results at the depinning transition ($\zeta_{n=2} >1$), 
it is clearly not a good starting point
to study the general phase diagram for the
dynamics at finite temperature and finite force.
On the contrary, the metric constraint is always a safe choice,
since by definition it cannot produce an unphysical
 roughness strictly larger than one.

\subsection{ Interaction between the segments of size $l^*(F,T)$
in the effective directed trap model } 

For the reasons explained in details above,
we thus consider that the { \it microscopic } interactions between monomers
are given by the metric constraint.
The question is then : in the effective directed trap model described 
in section \ref{interactingtrap},
what is the corresponding renormalized interactions
between two consecutive segments of typical length $l^*(F,T)$
representing local quasi-equilibrated metastable states ?
In the absence of more refined arguments, we feel that the metric constraint
is actually also an appropriate renormalized interaction, since
it prevents the appearance of an unphysical roughness $\zeta>1$
and it allows all possible physically acceptable values $\zeta \leq 1$
for the roughness.

\end{document}